\documentclass[twocolumn,twocolappendix]{aastex631}
\usepackage{hyperref}  % For referencing
\usepackage{multirow}  % For merging rows
\usepackage{amsmath}
\usepackage{float}

\usepackage{amsfonts} % optional
\usepackage{graphicx} % often included
\usepackage[T1]{fontenc}

\makeatletter
\renewenvironment{acknowledgments}
  {\section*{Acknowledgments}}
  {}
\makeatother

\begin{document}

\title{Protostellar Outflows Shed Light on the Dominant Close Companion Star Formation Pathways}

\author[0009-0009-5567-2107]{Ryan Sponzilli}
\affiliation{Department of Astronomy, University of Illinois, 1002 West Green St, Urbana, IL 61801, USA}
\correspondingauthor{Ryan Sponzilli}
\email{ryants4@illinois.edu}

\author[0000-0002-4540-6587]{Leslie W. Looney}
\affiliation{Department of Astronomy, University of Illinois, 1002 West Green St, Urbana, IL 61801, USA}

\author[0000-0002-6195-0152]{John J. Tobin}
\affiliation{National Radio Astronomy Observatory, 520 Edgemont Rd., Charlottesville, VA 22903 USA}

\author{Frankie J. Encalada}
\affiliation{Department of Astronomy, University of Illinois, 1002 West Green St, Urbana, IL 61801, USA}

\author[0000-0001-7387-3898]{Austen Fourkas}
\affiliation{Department of Astronomy, University of Illinois, 1002 West Green St, Urbana, IL 61801, USA}

%% listed alphabetically
\author[0000-0001-5653-7817]{Hector Arce}
\affiliation{Department of Astronomy, Yale University, P.O. Box 208101, New Haven, CT 06520, USA}

\author[0000-0002-5216-8062]{Erin Cox}
\affiliation{NSF-Simons AI Institute for the Sky (SkAI), 172 E. Chestnut St., Chicago, IL 60611, USA}
\affiliation{Center for Interdisciplinary Exploration and Research in Astrophysics (CIERA), 1800 Sherman Avenue, Evanston, IL 60201, USA}

\author[0000-0002-9289-2450]{James Di Francesco}
\affiliation{NRC Herzberg Astronomy and Astrophysics, 5071 West Saanich Road, Victoria, BC V9E 2E7, Canada}
\affiliation{Department of Physics and Astronomy, University of Victoria, Victoria, BC V8P 5C2, Canada}

\author[0000-0003-3682-854X]{Nicole Karnath}
\affiliation{Space Science Institute, 4765 Walnut Street, Suite B, Boulder, CO 80301, USA}

\author[0000-0002-7402-6487]{Zhi-Yun Li}
\affiliation{Department of Astronomy, University of Virginia, P. O. Box 400325, 530 McCormick Road, Charlottesville, VA 22904-4325, USA}

\author[0009-0001-6486-6909]{Nadia Murillo}
\affiliation{Instituto de Astronomía, Universidad Nacional Autónoma de México, AP106, Ensenada, CP 22830, B.C., Mexico}

\author[0000-0003-1252-9916]{Stella Offner}
\affiliation{Department of Astronomy, University of Texas at Austin, TX 78712, USA}

\author[0000-0001-7474-6874]{Sarah Sadavoy}
\affiliation{Department of Physics and Astronomy, York University, Toronto, Ontario M3J 1P3, Canada}

\author[0000-0002-0549-544X]{Rajeeb Sharma}
\affiliation{Niels Bohr Institute, University of Copenhagen, Jagtvej 155A, 2200 Copenhagen N., Denmark}

%% no comments recieved from these people
% \author[0000-0002-7506-5429]{Guillem Anglada}
% \affiliation{TBD}

% \author[0000-0002-5216-8062]{Erin Cox}
% \affiliation{TBD}

% \author[0000-0002-9289-2450]{James Di Francesco}
% \affiliation{TBD}

% \author[0000-0001-5253-1338]{Kaitlin Kratter}
% \affiliation{TBD}

% \author[0000-0001-7629-3573]{Tom Megeath}
% \affiliation{TBD}

% \author[0000-0002-6737-5267]{Mayra Osorio}
% \affiliation{TBD}

%% Note that the \and command from previous versions of AASTeX is now
%% depreciated in this version as it is no longer necessary. AASTeX 
%% automatically takes care of all commas and "and"s between authors names.

%% AASTeX 6.31 has the new \collaboration and \nocollaboration commands to
%% provide the collaboration status of a group of authors. These commands 
%% can be used either before or after the list of corresponding authors. The
%% argument for \collaboration is the collaboration identifier. Authors are
%% encouraged to surround collaboration identifiers with ()s. The 
%% \nocollaboration command takes no argument and exists to indicate that
%% the nearby authors are not part of surrounding collaborations.

%% Mark off the abstract in the ``abstract'' environment. 
\begin{abstract}
Understanding the formation pathway for close-companion protostars is central to unraveling the processes that govern stellar multiplicity and very early star formation. We analyze a large sample of 51 Class 0/I close‑companion protostellar systems, of which 38 show detectable outflows, yielding 42 measured outflows used in our analysis. We use ALMA observations of 11 systems in Perseus and 40 systems in Orion. These companions formed either directly at these small scales ($\lesssim$500 au separations) via disk fragmentation or at larger scales ($>$1000 au separations) via turbulent fragmentation followed by inward migration. Because of differences in formation mechanism, the former is expected to have preferentially aligned disks and outflows, whereas the latter is expected to show no preferred alignment. The relative prevalence of these formation pathways remains uncertain, yet it is critical to forming a comprehensive picture of star formation.  We examine the distribution of position angles of companion protostars relative to the position angles of their molecular outflows. The outflow, as traced by $^{12}$CO ($J=2\rightarrow1$), is a useful proxy for the angular momentum of the system, expected to be orthogonal to the binary orbital plane. We use a simple model to account for a random sampling of inclination and orbital phase in each system, finding that the observations are consistent with a distribution where the outflows are preferentially orthogonal to the companions. Based on this analysis, we suggest disk fragmentation is the dominant formation pathway for close-companion protostellar systems.
\end{abstract}

%% Keywords should appear after the \end{abstract} command. 
%% The AAS Journals now uses Unified Astronomy Thesaurus concepts:
%% https://astrothesaurus.org
%% You will be asked to selected these concepts during the submission process
%% but this old "keyword" functionality is maintained in case authors want
%% to include these concepts in their preprints.
%\keywords{Classical Novae (251) --- Ultraviolet astronomy(1736) --- History of astronomy(1868) --- Interdisciplinary astronomy(804)}

%% From the front matter, we move on to the body of the paper.
%% Sections are demarcated by \section and \subsection, respectively.
%% Observe the use of the LaTeX \label
%% command after the \subsection to give a symbolic KEY to the
%% subsection for cross-referencing in a \ref command.
%% You can use LaTeX's \ref and \label commands to keep track of
%% cross-references to sections, equations, tables, and figures.
%% That way, if you change the order of any elements, LaTeX will
%% automatically renumber them.
%%
%% We recommend that authors also use the natbib \citep
%% and \citet commands to identify citations.  The citations are
%% tied to the reference list via symbolic KEYs. The KEY corresponds
%% to the KEY in the \bibitem in the reference list below. 

\section{Introduction} 
\label{sec:intro}
Stars are born within molecular clouds through gravitational collapse, which happens when internal sources of support (thermal, magnetic, turbulence, etc.) are weaker than self-gravity \citep{Shu_et_al_1987, mckee&ostriker_2007}. Gravitational collapse occurs on multiple scales within molecular clouds. It proceeds from clouds, to clumps, to cores, to protostars. During the collapse of cores to protostars, denser gas collapses more rapidly causing the denser interiors of cores to collapse first. This inside-out collapse and the conservation of angular momentum results in a central, flattened, and rotating disk embedded in a broader, less dense, and infalling envelope \citep{Adams_et_al_1987}.

Star formation not only involves the infall dynamics of gravitational collapse but also outflow phenomena in the form of jets, winds, and molecular outflows. Jets and winds drive molecular outflows by shocking and entraining ambient material. Highly collimated jets with velocities of 100-1000 km~s$^{-1}$ are believed to be launched from protostars through dynamical interactions of accreted material with magnetic fields \citep[e.g.,][]{Frank_2014, Bally_2016}. By contrast, disk winds contributing to the lower-velocity outflow launch material from various radii on a rotating disk \citep[e.g.,][]{Feeney-Johansson_et_al_2024}, and may be ejecting material directly from the disk surface \citep{pascucci_2023}. Molecular outflows and jets are typically launched perpendicular to the major disk axis \citep[e.g.,][]{Hsieh_2023, Lee_et_al_2016}, and outflows often have clearly defined edges of parabolic or conical shapes on one or both sides of the protostar, resulting in typical v-shaped morphologies. These outflows can be used as a proxy for the angular momentum of a system \citep[e.g.,][]{Offner_et_al_2016}.

It is well established that approximately half of all solar-type stars reside in binary or higher-order systems \citep[e.g.,][]{Raghavan_et_al_2010, Offner_et_al_2023}. Young stellar objects (YSOs) have even higher rates of multiplicity \citep[e.g.,][]{Looney_et_al_2000, Tobin_et_al_2022}. This statistic implies that star formation is inherently a process that, more often than not, creates more than one star. The specifics of how this occurs and how such systems evolve are active areas of research \citep[][and references therein]{Offner_et_al_2023}.

The distribution of binary separations appears bimodal with one peak at scales $\sim$100 au and a second peak at scales $>$1000 au \citep{Tobin_et_al_2016, Tobin_et_al_2022}. This bimodal distribution suggests two distinct formation mechanisms acting at different scales, independently producing close and wide companion systems. These two formation mechanisms are believed to be disk fragmentation \citep{Tohline_2002, Kratter&Lodado_2016}, and turbulent fragmentation of the core \citep{Tohline_2002, Offner_et_al_2010, Offner_et_al_2023}, respectively. Fragmentation is the process by which self-gravitating objects develop substructures that evolve and collapse independently \citep{Offner_et_al_2023}.

Briefly, turbulent fragmentation occurs when turbulence produces density perturbations that become gravitationally unstable and collapse, whereas disk fragmentation occurs when the disk around a YSO becomes gravitationally unstable and fragments to form a second protostar at close separations \citep[e.g.,][]{Kratter_et_al_2010, Offner_et_al_2010, Offner_et_al_2016}. While the former method preferentially produces wide-separation protostellar systems on $>$ 1000 au scales, dynamical interactions can cause the migration of these companions to close separations consistent with disk fragmentation \citep[e.g.,][]{Offner_et_al_2010, Bate_2012}. Thus, it is an open question whether close multiple systems were formed in situ from disk fragmentation or via turbulent fragmentation followed by inward migration. The alignment of the angular momentum directions of companions, as probed by outflows, is one proposed means of distinguishing between close-companions formed from disk fragmentation or from turbulent fragmentation followed by inward migration. 

In this paper, we seek to differentiate between these two mechanisms using angular momentum as a discriminant; close binaries formed via disk fragmentation should have an angular momentum orthogonal to the binary's orbit, whereas turbulent fragmentation would produce more random alignments. We use outflow position angles (PAs) as a proxy for the angular momentum directions of the protostars and the binary separation PA to define the orientation of the major disk axis. We measure protostellar outflow PAs in Perseus and Orion on new data from the Atacama Large Millimeter/Submillimeter Array (ALMA). 

In Section \ref{sec:obs}, we present the observations. In Section \ref{sec:results}, we describe how we measured outflow PAs, examine the outflows and their measurements, and compare the observed PA distribution to a range of simulated models. In Section \ref{sec:dis}, we discuss how various close-companion systems appear consistent or inconsistent with the two formation pathways: turbulent fragmentation and disk fragmentation. In Section \ref{sec:conc}, we present our conclusions.

\section{Observations} \label{sec:obs}

\subsection{Sample Selection} \label{sec:obs:sample-selection}
This study focuses on Class 0 and Class I YSOs because they represent the most pristine systems and exhibit prominent outflows that can be used to infer angular momentum. Class 0 sources are low-mass protostars that have not yet accreted the majority of their stellar massand remain deeply embedded within a large, infalling envelope. Class I sources are slightly more evolved, having accreted more material from the envelope \citep{Lada&Wilking_1984, Adams_et_al_1987, Greene_et_al_1994, Andre_et_al_1993}.

The NSF's Karl G. Jansky Very Large Array (VLA) conducted the VLA/ALMA Nascent Disk and Multiplicity Survey (VANDAM) in both Orion and Perseus \citep{Tobin_et_al_2016, Tobin_2020}. Combining the VANDAM Orion survey with the complete ALMA Cycle 3/4 survey of Orion’s protostellar population, we identified 40 close-companion protostars. In this paper, we define close companions as protostars with projected separations $\lesssim$500 au \citep{Tobin_et_al_2022}, although the largest separation in our sample is 573 au due to older distance estimates. Similarly, from the VANDAM Perseus survey and a previous ALMA Cycle 2 program \citep{Tobin_et_al_2018}, we identified 11 close-companion protostars. While our focus is on close binaries, the sample includes some higher-order multiples; for these, we focus on only the innermost close pair.

These 51 systems were observed with ALMA as part of two separate project codes (one for Perseus and one for Orion) during cycles five and six. In addition to these 51, Per-emb-5 was observed; however it is now known not to be a binary system so we do not consider it in this paper \citep{Reynolds_et_al_2024}. The full set of ALMA observations from these projects has not yet been published. From the full sample, we find that 38 systems (with 42 outflows) are the most suitable for the detailed analysis in this study, see Section \ref{sec:results:measurments} for more details. In each field, ALMA observed the dust continuum and multiple molecular emission lines. We focus only on the continuum and $^{12}$CO ($J=2\rightarrow1$) molecular line data since $^{12}$CO ($J=2\rightarrow1$) emission is a well-established tracer of molecular outflows \citep[e.g.,][]{Bally&Lada1983, Bontemps1996}. The observations of these 51 systems are detailed below. Distances to systems in Orion are adopted from \cite{Tobin_et_al_2022}, and distances to systems in Perseus are adopted as an average of 300 pc \citep{Zucker+2018}. Basic details of outflows measured and each field observed are provided in Tables \ref{table:by_ouflow} and \ref{table:by_field}.

\subsection{Orion Data} \label{sec:obs:orion}
The Orion multiple systems were observed with the ALMA main array (12~m), project code: 2018.1.01038.S. The observations were in Band 6 at 1.3~mm. The sources were observed in two different groups: Orion A and Orion B. 

In Orion A, we observed in nine execution blocks (EBs), with approximately 20 minutes on each source, on 2018 October 11 (twice; semi-pass), 2018 October 14 (three times), 2018 October 16 (twice), and 2018 October 19 (twice) with $\sim$45 antennas operating and sampling baselines between $\sim$15~m and $\sim$2500~m. The mean precipitable water vapor (pwv) was $\sim$1~mm. The semi-pass data was reduced and included.

In Orion B, we observed in eight EBs, with approximately 20 minutes on each source, on 2018 October 2 (semi-pass), 2018 October 3 (semi-pass), 2018 October 9 (semi-pass), 2018 October 26, 2018 October 28 (three times), and 2018 November 23 with $\sim48$ antennas. The mean pwv was $\sim1$~mm. The semi-pass data was reduced and included.

The correlator was configured for the first baseband to observe two $\sim$59~MHz spectral windows with 30.5~kHz channels centered on $^{13}$CO ($J=2\rightarrow1$) and C$^{18}$O ($J=2\rightarrow1$). Then the second baseband was configured with four 59~MHz spectral windows with 122~kHz channels centered on H$_2$CO ($J=3_{03}\rightarrow2_{02}$), ($J=3_{22}\rightarrow2_{21}$),  ($J=3_{21}\rightarrow2_{20}$), and SO ($J_N = 6_5\rightarrow5_4$). The third baseband was configured with one 938~MHz spectral window with 488~kHz channels centered on $^{12}$CO ($J=2\rightarrow1$). The fourth baseband was configured to observe a 1.875~GHz continuum band centered at 233.5~GHz with 1920 977~kHz channels.

The raw visibility data were reduced using the ALMA pipeline integrated within CASA 5.6.1 \citep{CASA}. We additionally performed self-calibration with CASA 5.6.1 on the continuum data to increase the signal to noise ratio. We performed self-calibration of the continuum data on all sources using two rounds of phase-only self-calibration where the first solution interval corresponded to the length of a single scan and the second solution interval corresponded to 18.15~s or 6.05~s; 6.05~s corresponds to the length of a single integration. Following phase-only self-calibration, we also applied amplitude self-calibration, using a solution interval that was the length of a scan, and we used the option \texttt{solnorm=True} to normalize the amplitude solutions to 1, rather than allowing them to scale freely, which prevents the flux density from being significantly altered erroneously. Following the completion of self-calibration on the continuum, the solutions were also applied to the spectral line data. We used Briggs weighting with a \texttt{robust=0.5} with calls to \texttt{tclean} for both the dust continuum and the spectral lines. The continuum images were produced using standard calls to the ALMA pipeline tasks \texttt{hif\_makeimlist} and \texttt{hif\_makeimages}. We used options of \texttt{hif\_makeimages} and \texttt{hif\_editimlist} to image the spectral windows with the desired channel binning and frequency range. For $^{12}$CO ($J=2\rightarrow1$), we used 1~km~s$^{-1}$ channels and imaged $\pm$100~km~s$^{-1}$ of the source velocity. 

The $^{12}$CO ($J=2\rightarrow1$) cubes typically had 1372$\times$1372 pixels, a pixel scale of 0\farcs034, and a synthesized beam of 0\farcs22$\times$0\farcs17. The continuum images were typically 1280$\times$1280~pixels and a pixel scale of 0\farcs031, and the synthesized beam was 0\farcs21$\times$0\farcs15. The noise in the 1.3~mm continuum was $\sim$0.2~mJy~beam$^{-1}$, and the noise in the $^{12}$CO ($J=2\rightarrow1$) cubes was $\sim$2~mJy~beam$^{-1}$.

\subsection{Perseus Data} \label{sec:obs:perseus}
The Perseus multiple systems were observed with the ALMA main array (12~m), project code: 2017.1.00053.S. The observations were in Band 6 at 1.3~mm. The sources were observed in five EBs, with approximately 19 minutes on each source, on 2017 December 17, 2018 January 7, 2018 January 11, 2018 January 19, and 2018 September 20 with 43 to 45 antennas operating and sampling baselines between $\sim$15~m and $\sim$2500~m. The observations were executed within a $\sim$1.1~hour block, and the total time spent on each source was $\sim$17.6 to 20.2~minutes. The pwv ranged over 0.54 to 2.57~mm during the different executions.

The correlator was configured for the first baseband to observe a 2 GHz continuum band centered at 232.5~GHz and observed in TDM mode with 128 channels. Then the second baseband was configured with two 234~MHz spectral windows with 488 kHz channels centered on $^{12}$CO ($J=2\rightarrow1$) and N$_2$D$^{+}$ ($J=3\rightarrow2$) and $^{13}$CS ($J=5\rightarrow4$). The third baseband was configured with four 59~MHz spectral windows with 122~kHz channels centered on H$_2$CO ($J=3_{03}\rightarrow2_{02}$), ($J=3_{22}\rightarrow2_{21}$),  ($J=3_{21}\rightarrow2_{20}$), and SO ($J_N = 6_5\rightarrow5_4$). The fourth baseband was configured with two 59~MHz spectral windows with 61~kHz channels centered on $^{13}$CO ($J=2\rightarrow1$) and C$^{18}$O ($J=2\rightarrow1$).

The raw visibility data were reduced by the ALMA pipeline integrated within CASA 5.4.0 \citep{CASA}. We additionally performed self-calibration on the continuum data to increase the signal-to-noise ratio. We performed self-calibration of the continuum data on all sources also using CASA 5.6.1, using two rounds of phase-only self-calibration, where the first solution interval corresponded to the length of a single scan and the second solution interval corresponded to 12.1~s or 6.05~s; 6.05~s corresponds to the length of a single integration. Again, afterwards we also applied amplitude self-calibration using a solution interval that was the length of a scan with the \texttt{solnorm=True} option. Following the completion of self-calibration on the continuum, the solutions were also applied to the spectral line data. The data were imaged using a custom recipe of the ALMA imaging pipeline in CASA version 5.4.0, using Briggs weighting with \texttt{robust=0.5} for all images. The continuum images were produced using standard calls to the ALMA pipeline tasks \texttt{hif\_makeimlist} and \texttt{hif\_makeimages}. We used options of \texttt{hif\_makeimages} and \texttt{hif\_editimlist} to image the spectral windows with the desired channel binning and frequency range. For $^{12}$CO ($J=2\rightarrow1$) we used 0.66~km~s$^{-1}$ channels and imaged $\pm$100~km~s$^{-1}$ of the source velocity. 

The $^{12}$CO ($J=2\rightarrow1$) cubes typically had 1000$\times$1000~pixels, a pixel scale of 0\farcs041, and a synthesized beam of 0\farcs29$\times$0\farcs21. The continuum images were typically 1372$\times$1372 pixels, a pixel scale of 0\farcs03, and the synthesized beam was 0\farcs28$\times$0\farcs21. The noise in the 1.3~mm continuum was $\sim$0.5~mJy~beam$^{-1}$, and the noise in the $^{12}$CO ($J=2\rightarrow1$) cubes was 3.7~mJy~beam$^{-1}$.

The source L1448 IRS3C, was not self-calibrated in the same way as the others because it had been unintentionally left out of the original processing. Instead of manually processing the data, we imaged those data with the ALMA pipeline included in CASA 6.6.1 using automated self-calibration through the imaging service available on the NRAO archive. Comparison of self-calibration done through the automated imaging service versus the manual self-calibration showed very similar results, and the ALMA imaging pipeline was executed in a very similar way to all the other images as well.

\section{Results} \label{sec:results}

\subsection{Measuring Outflows} \label{sec:results:methods}
We define a single outflow as a pair of redshifted and blueshifted lobes, although sometimes only one lobe is detected. We refer to the redshifted side of the outflow as the red lobe and/or red channels and the blueshifted side as the blue lobe and/or blue channels. We describe outflow shapes as being v-shaped or u-shaped. A v-shape is composed of two straight lines like a cone, while a u-shaped is composed of two curved lines like a parabola. We refer to these lines as the edges of the outflow, and the space between them as the cavity. 

We identify, where possible, which source in a system launches the outflow, and the observations fall into three categories: the outflow is clearly coming from one particular source; the outflow is coming from in between two sources and/or they both appear to feed the outflow; or the sources are too close together to resolve the outflow source. The determinations are made by eye using the $^{12}$CO ($J=2\rightarrow1$) data cubes for the outflow positions and the continuum data for the source positions. 

We measure outflow PAs by eye with a protractor on the $^{12}$CO ($J=2\rightarrow1$) data within Cube Analysis and Rendering Tool for Astronomy \citep[CARTA;][]{carta}. First, we load the $^{12}$CO ($J=2\rightarrow1$) and 1.3~mm (2 GHz band centered at 232.5 GHz) continuum data into CARTA. Then we annotate the source positions from the continuum peak as a reference for determining the launching source of the outflow. Next, we locate the rough velocity range of the red lobe in the $^{12}$CO ($J=2\rightarrow1$) data. Then we use the line tool in CARTA to trace both edges of the v-shape, such that the lines intersect the source position in the continuum data. If the source exhibits a u-shaped outflow, we trace the outflow edge near its base. Subsequently, we also note whether the outflow shape is clearly launching from one source or between the two sources. 

We trace the outflow edges in a velocity channel that has the clearest shape, while also making sure the lines are consistent with the outflow edges in the neighboring velocity channels. While the outflow may narrow or widen with velocity, such variations should have only a minor impact on the center PA (assuming the width is approximately symmetric), so the choice of channel for measuring the edges is not expected to strongly affect the result. Then we overlay a protractor application and measure the PAs of both edges of the lobe counterclockwise from north (PA = 0$^\circ$). Then we determine the PA of the center of the outflow by adding half the difference between the two edges to the first edge.

We repeat this process for the blue lobe, independently of the red lobe. If an outflow has two lobes, we measure each separately and report the lobe-averaged PA and the lobe PA offset (the discrepancy between the red and blue lobe PAs). If only one lobe was detected, only the PA of that lobe is used. The PAs are specified such that they always refer to the direction of the blue lobe. 

In certain fields, we encountered a system with two outflows. These systems fall under two cases: binaries where each companion drives its own outflow, and hierarchical triples where the inner binary drives one outflow and the tertiary companion drives another. We measure each outflow independently.

For each outflow PA measurement, we classify it as either high confidence (HC) or low confidence (LC). We adopt uncertainties of 10$^\circ$ for HC outflows and 30$^\circ$ for LC outflows. Factors that make a measurement LC include: a large PA offset between outflow lobes ($\gtrsim$20$^\circ$), low signal to noise ratio, and extremely wide outflow cavities ($\sim$180$^\circ$). These factors make it difficult to measure a PA accurately, so we adopt a higher uncertainty for LC outflows. Specific factors for each outflow are described in \ref{sec:appendix:fields-with-emission}. 

Each Perseus system has previously been examined and had its outflow PAs measured \citep[e.g.,][]{Plunkett_2013, Lee_et_al_2016, Stephens_et_al_2017, Dunham_et_al_2024}. In this analysis, we independently remeasured outflows in each Perseus system and found general agreement with previous results, with a few noted discrepancies discussed in \ref{sec:appendix:fields-with-emission}.

We calculate the PA of the binary separation (the projected binary axis) and compare it to the outflow PA. In the case of higher-order multiple systems, we consider the innermost binary. The binary separation PAs are derived using the  coordinates from each YSO reported by \cite{Tobin_et_al_2018, Tobin_et_al_2022} and \cite{Reynolds_et_al_2024}. For each outflow, we then calculate $\Delta$PA, defined as the difference between the outflow PA and the binary separation PA. A $\Delta$PA value near 90$^\circ$ indicates an outflow that is orthogonal to the binary separation PA, at least in projection. We do not calculate $\Delta$PA for hierarchical triples with multiple outflows, since comparing the tertiary outflow PA to the inner binary separation PA is not physically meaningful.

A summary of the measured outflows is presented in Table \ref{table:by_ouflow}, and the complete set of measured PAs are given in Table \ref{table:by_outflow_full}. Systems with two outflows have an entry for each outflow in the tables. Figures \ref{fig:fig_1} and \ref{fig:fig_2} show examples of the outflows and the measured position angles.

\subsection{Outflow Detections and Morphologies} \label{sec:results:measurments}

\begin{figure*}[ht!]
\centering
\includegraphics[width=0.98\textwidth]{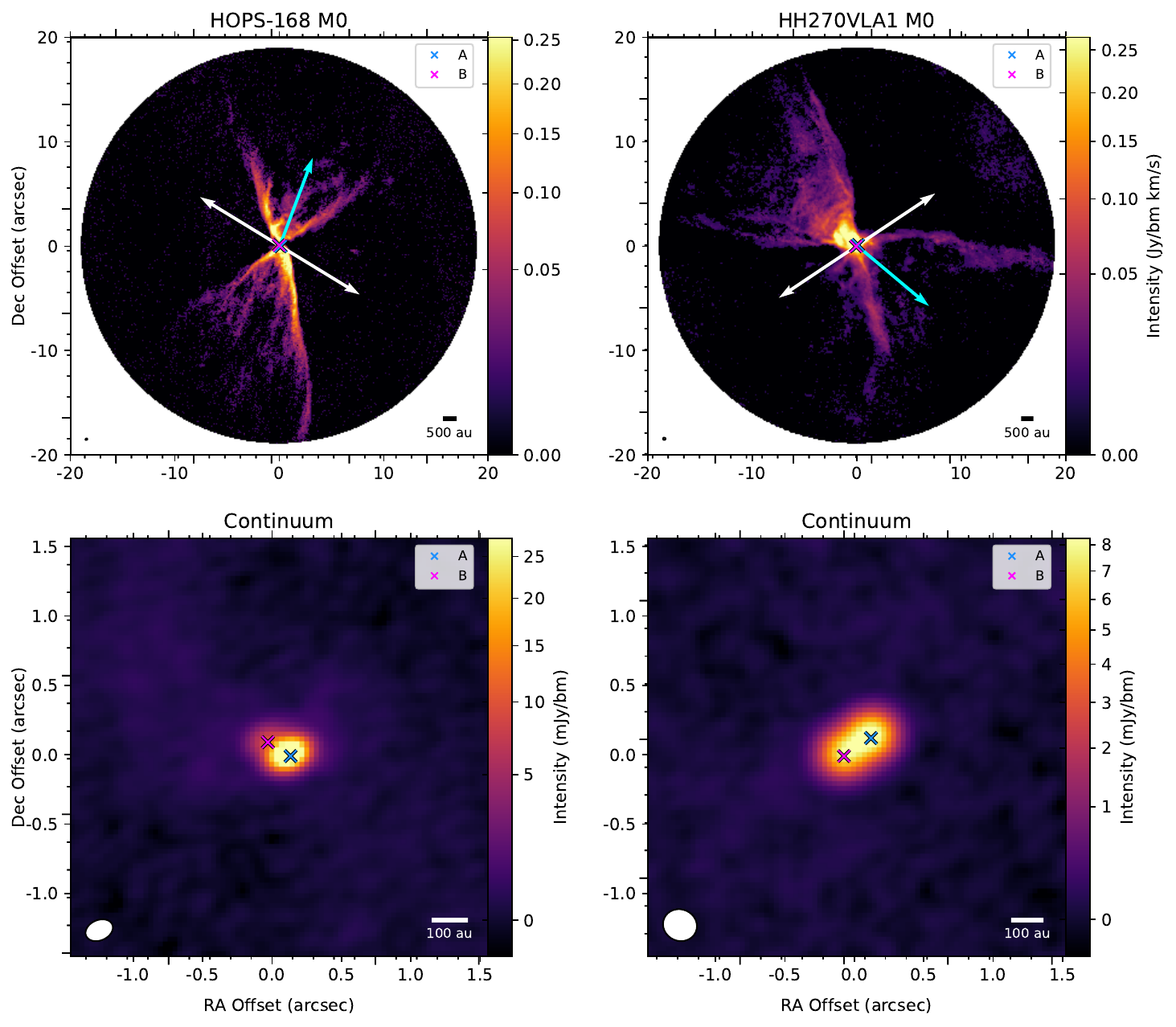}
\caption{A subset of two typical close binary systems with one outflow. These are integrated intensity (moment 0) maps of $^{12}$CO ($J=2\rightarrow1$). The axes are relative to the center of the image which is given in Table \ref{table:by_field} along with the integrated velocities. Below each is a zoom-in panel of the continuum (2 GHz band centered at 232.5 GHz) sources, with a box size of 3" by 3". The white arrows indicate the binary separation PA, and the cyan arrow(s) indicates the measured outflow PA(s) in the direction the blue-shifted lobe. All images are shown in Appendix \ref{sec:appendix}.}
\label{fig:fig_1}
\end{figure*}

\begin{figure*}[ht!]
\centering
\includegraphics[width=0.98\textwidth]{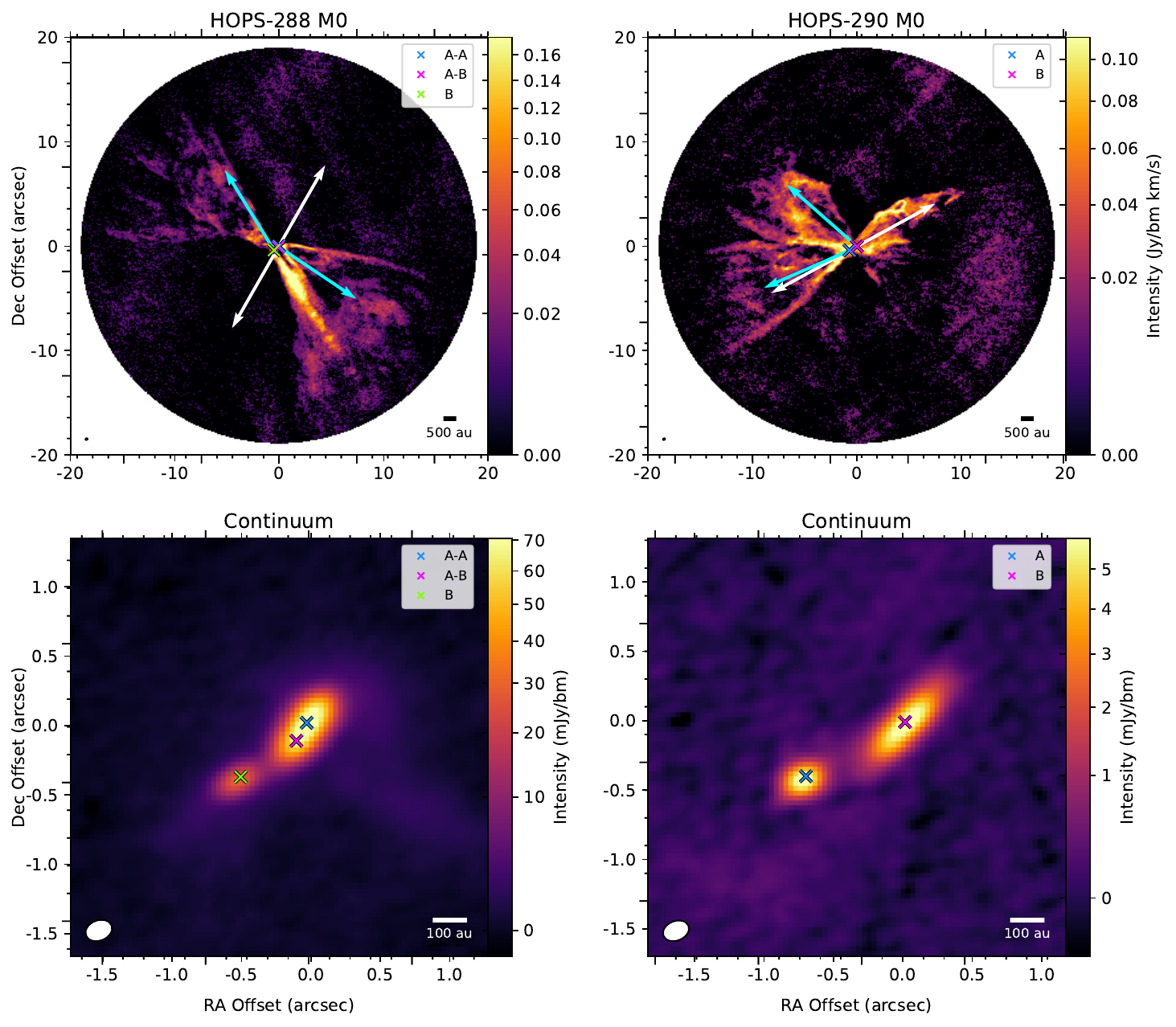}
\caption{A subset of two close-multiple systems with more complex outflow phenomena. These are integrated intensity (moment 0) maps of $^{12}$CO ($J=2\rightarrow1$). The axes are relative to the center of the image which is given in Table \ref{table:by_field} along with the integrated velocities. Below each is a zoom-in panel of the continuum (2 GHz band centered at 232.5 GHz) sources, with a box size of 3" by 3". The white arrows indicate the binary separation PA, and the cyan arrow(s) indicates the measured outflow PA(s) in the direction the blue-shifted lobe. All images are shown in Appendix \ref{sec:appendix}. We categorize HOPS-288 (left) as a hierarchical triple with two outflows. We categorize HOPS-290 (right) as a binary with two outflows.}
\label{fig:fig_2}
\end{figure*}

\begin{figure}[ht!]
\centering
\includegraphics[width=0.48\textwidth]{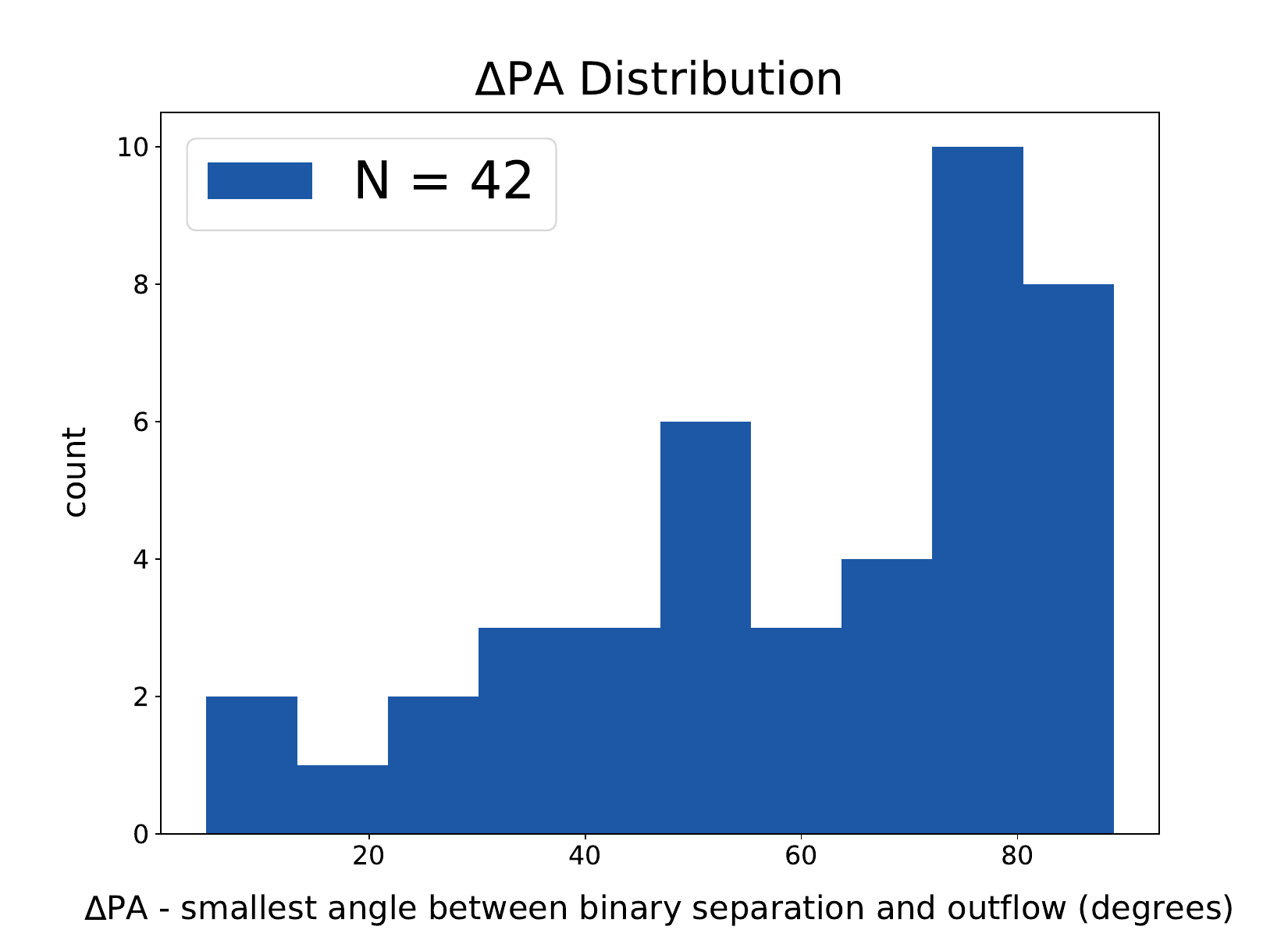}
\caption{The observed distribution of $\Delta$PA values. The distribution includes 42 values for $\Delta$PA: 34 single outflow binaries and 4 where each binary component drives its own outflow.}
\label{fig:histogram}
\end{figure}

All targets were detected in $^{12}$CO ($J=2\rightarrow1$) emission. The Perseus source positions were slightly offset from those reported in \cite{Tobin_et_al_2018}, likely due to proper motion and/or observational effects arising from differences in spatial resolution. We therefore applied manual positional corrections to align the targets with the continuum peak; the resulting offsets are listed in Table~\ref{table:offsets}.

We measured at least one outflow PA in 27 of 40 close-companion systems in Orion and in all 11 systems in Perseus. Of the remaining 13 Orion systems, seven exhibited complex or ambiguous structures that prevented a reliable PA measurement, and six showed no extended $^{12}$CO ($J=2\rightarrow1$) emission (see Section~\ref{sec:appendix:fields-without-emission}). For these 13 systems, we were unable to determine an outflow PA. In total, we measure at least one outflow in 38 systems.

We note a few interesting cases. We observe four binary systems with a single outflow that appears to be launched exclusively by one source (Per-emb-2, HOPS-32, HOPS-395, HOPS-400). We observe four binary systems with two outflows, where each companion drives its own outflow (HOPS-290, Per-emb-12, Per-emb-27, and Per-emb-35). We observe five hierarchical triple systems where the inner binary drives one outflow and the tertiary companion drives another (HOPS-12, HOPS-92, HOPS-203, HOPS-288, Per-emb-33). In our numerical analysis we only consider the outflow PA of the inner binary and disregard the outflow PA of the tertiary companion. 

Figure \ref{fig:fig_1} illustrates two representative systems with a single outflow. The edges of the molecular outflow are seen on either side of each system. They display a characteristic v-shape with an apparent cavity that points back to their origin near the continuum sources that denote the protostar locations. Figure \ref{fig:fig_2} shows two examples of systems with multiple outflows. Nearly all of our observed outflow morphologies appear consistent with being driven by a slow, wide-angle wind characterized by small scales ($<$2000 au), striking conical or parabolic shapes, semi-opening angles of 10$^\circ$-40$^\circ$, and an inner low-brightness cavity \citep[e.g.,][]{pascucci_2023, Feeney-Johansson_et_al_2024}.

The final analysis focuses on 38 systems for which outflow position angles could be reliably determined. This leaves us with 42 values for $\Delta$PA: 34 single outflow binaries and 4 where each binary component drives its own outflow (8 PAs, 2 per binary system). Among all 38 systems with measured outflow PAs, the maximum binary separation is 573 au, the minimum is 24 au, and the median is 100 au. As shown in Figure \ref{fig:histogram}, the observed $\Delta$PA values are strongly skewed toward large angles (70°–90°), indicating that most outflows are nearly orthogonal to the binary separation axis in projection.

\subsection{Model Comparisons} \label{sec:results:models}
To interpret the observed distribution of $\Delta$PA values, we must consider that any individual value of $\Delta$PA $\not\approx$ 90$^\circ$ may be a projection effect of inclination instead of a misaligned outflow, and even $\Delta$PA $\approx$ 90$^\circ$ could be a projection effect. To account for this ambiguity, we generate 101 models ranging from 100\% preferentially aligned ($\Delta$PA is orthogonal in 3D) to 0\% preferentially aligned ($\Delta$PA is random in 3D), where each model incremented by 1\%. These boundary distributions are hereafter referred to as the orthogonal distribution and random distribution, and the 99 distributions in between them are referred to by their percentage of orthogonal $\Delta$PAs (e.g., 42\% orthogonal).

The 101 distributions were constructed with Monte Carlo-style simulations, each based on 100,000 random draws. For the orthogonal distribution, a binary separation PA was sampled from a uniform random distribution between 0$^\circ$ and 90$^\circ$, and the inclination was similarly randomly sampled between 0$^\circ$ and 90$^\circ$, but with a uniform distribution in $\cos i$, yielding a mean inclination of 60$^\circ$. The binary separation PA was then projected onto the 2-D plane using the inclination. The expected orthogonal outflow position angles were found by subtracting projected separation angles from 90$^\circ$. For the random distribution, we used the same procedure to generate binary separation PAs and inclinations, followed by projection. We then assigned a random outflow PA offset drawn uniformly between 0$^\circ$ and 90$^\circ$. Subtracting the projected separation angle from this random offset yields the simulated random distribution. The 99 distributions between the orthogonal and random distributions are created by concatenating the corresponding percentages of these two boundary distributions together.

For each of these simulated distributions, we compare them to our observed distribution of $\Delta$PA with a 2-sample Kolmogorov-Smirnov (KS) test. The KS test compares the cumulative distributions of two datasets to assess whether they are likely drawn from the same underlying distribution. The p-value quantifies the probability of observing the data if the null hypothesis (the distributions are the same) is true, with higher values indicating greater statistical consistency. We use these p-values to identify the fraction of orthogonal outflows for which the data is no longer consistent with the observations. Using a significance level of p = 0.05, we fail to reject the null hypothesis of a dominantly orthogonal model; at this threshold, our data are consistent with a $>$71\% orthogonal model. The highest KS-test p-value (p $\sim$0.68) occurs for a 94\% orthogonal distribution, indicating this model provides the closest match to the data. We caution, however, that these p-values reflect only statistical consistency with, rather than proof of, the null model.

\begin{figure}[ht!]
\centering
\includegraphics[width=0.48\textwidth]{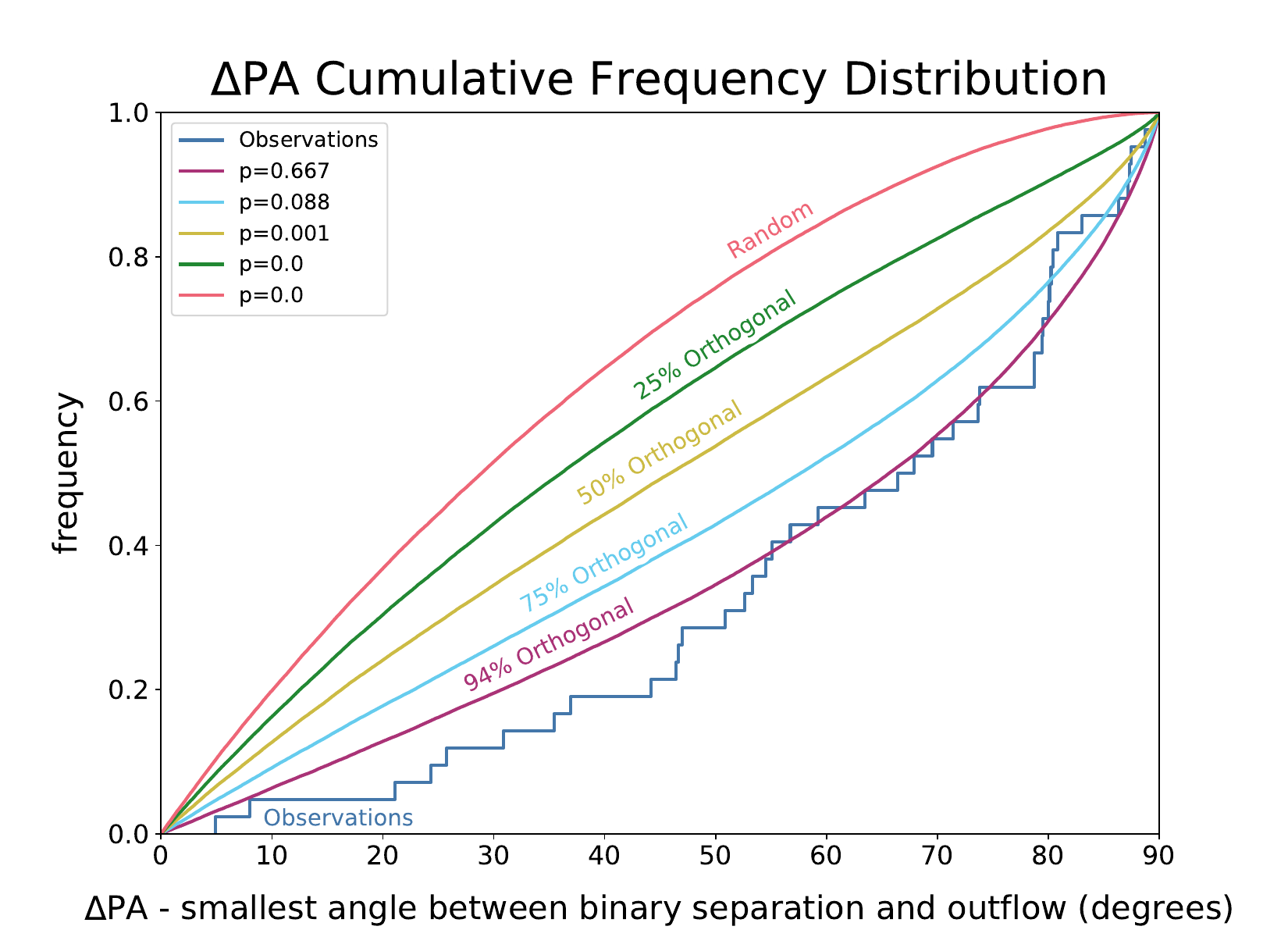}
\caption{The cumulative frequency distribution of $\Delta$PA. We plot the observed data along with the best fit model and four other models for reference. The observations most closely match the simulated distribution of 94\% orthogonal outflows.}
\label{fig:cumulat}
\end{figure}

Figure \ref{fig:cumulat} provides a snapshot of these models and associated p-values. We plot the observed data along with the best fit model (94\%) and four other models for reference. The $\Delta$PA observations most closely match the simulated distribution of 94\% orthogonal outflows. This model is for the combined $\Delta$PA of both Perseus and Orion. There are no large differences between the two samples, but the lower number of Perseus samples places less constraint on its outflow distribution.

% \clearpage
% \startlongtable
\begin{deluxetable*}{lllllllll}
\tablenum{1}
\tablecaption{Outflow Summary}
\tablehead{
    \colhead{Field} & 
    \colhead{Type\tablenotemark{a}} & 
    \colhead{Sep\tablenotemark{b, c}\ (au)} & 
    \colhead{Class\tablenotemark{c}} &
    \colhead{Source\tablenotemark{d}} &
    \colhead{Conf.\tablenotemark{e}} & 
    \colhead{Outflow PA\ ($^\circ$)\tablenotemark{f}} & 
    \colhead{Binary PA\ ($^\circ$)} & 
    \colhead{$\Delta$PA\ ($^\circ$)}
}
\startdata
Per-emb-22 & Binary & 225.4 & 0 & A+B & HC & 309 & 265 & 44 \\
L1448 IRS3C & Binary & 75.3 & 0 & A+B & HC & 307 & 11 & 63 \\
Per-emb-33 & Binary & 79.6 & 0 & A+B & HC & 303 & 329 & 26 \\
Per-emb-33 & Tertiary & 263.2 & 0 & C & HC & 288 & \nodata & \nodata \\
Per-emb-17 & Binary & 83.4 & 0 & A+B & HC & 236 & 147 & 89 \\
Per-emb-35 & Binary & 572.5 & I & A & HC & 119 & 64 & 55 \\
Per-emb-35 & Binary & 572.5 & I & B & HC & 310 & 64 & 66 \\
Per-emb-27 & Binary & 186.0 & 0 & A & HC & 203 & 195 & 8 \\
Per-emb-27 & Binary & 186.0 & 0 & B & HC & 294 & 195 & 81 \\
Per-emb-36 & Binary & 93.3 & I & A+B & LC & 226 & 0 & 46 \\
Per-emb-44 & Binary & 90.1 & 0 & A+B & LC & 139 & 268 & 51 \\
Per-emb-12 & Binary & 549.0 & 0 & B & HC & 208 & 308 & 80 \\
Per-emb-12 & Binary & 549.0 & 0 & A & HC & 180 & 308 & 53 \\
Per-emb-18 & Binary & 25.5 & 0 & A+B & HC & 167 & 259 & 87 \\
Per-emb-2 & Binary & 24.1 & 0 & B & HC & 126 & 327 & 21 \\
HOPS-32 & Binary & 162.2 & 0 & B & HC & 339 & 233 & 74 \\
HOPS-12 & Tertiary & 1853.5 & 0 & A & HC & 337 & \nodata & \nodata \\
HOPS-12 & Binary & 80.6 & 0 & B-A & HC & 337 & 45 & 68 \\
HOPS-92 & Tertiary & 498.8 & FS & B & HC & 89 & \nodata & \nodata \\
HOPS-92 & Binary & 108.4 & FS & A-A+A-B & HC & 256 & 209 & 47 \\
HOPS-84 & Binary & 275.5 & I & A+B & HC & 263 & 209 & 53 \\
HOPS-75 & Binary & 99.8 & 0 & A+B & LC & 280 & 187 & 87 \\
HOPS-182 & Binary & 381.5 & 0 & A+B & HC & 55 & 198 & 37 \\
HOPS-168 & Binary & 73.7 & 0 & A+B & HC & 339 & 59 & 79 \\
HOPS-203 & Binary & 49.7 & 0 & A+B & HC & 324 & 201 & 57 \\
HOPS-203 & Tertiary & 1085.5 & 0 & C & HC & 281 & \nodata & \nodata \\
HOPS-173 & Binary & 103.0 & 0 & A+B & LC & 67 & 146 & 79 \\
HOPS-193 & Binary & 100.0 & I & A+B & HC & 274 & 177 & 83 \\
HOPS-158 & Binary & 46.0 & FS & A+B & LC & 56 & 115 & 59 \\
HOPS-395 & Binary & 65.4 & 0 & B & HC & 174 & 263 & 89 \\
HOPS-288 & Binary & 61.3 & 0 & A-A+A-B & HC & 236 & 150 & 86 \\
HOPS-288 & Tertiary & 221.9 & 0 & B & HC & 32 & \nodata & \nodata \\
HOPS-290 & Binary & 332.7 & 0 & A & HC & 114 & 298 & 5 \\
HOPS-290 & Binary & 332.7 & 0 & B & HC & 49 & 298 & 70 \\
HOPS-281 & Binary & 311.6 & FS & A+B & HC & 206 & 253 & 47 \\
HOPS-282 & Binary & 70.1 & I & A+B & HC & 3 & 263 & 80 \\
HOPS-384 & Binary & 49.0 & 0 & A+A-B & LC & 4 & 40 & 35 \\
HOPS-304 & Binary & 272.7 & FS & A+B & LC & 152 & 252 & 80 \\
HOPS-400 & Binary & 184.0 & 0 & A & HC & 87 & 158 & 71 \\
HOPS-213 & Binary & 114.4 & FS & A+B & LC & 84 & 319 & 55 \\
HOPS-312 & Binary & 134.7 & 0 & A+B & HC & 36 & 137 & 80 \\
HOPS-363 & Binary & 79.5 & FS & A+B & LC & 273 & 124 & 31 \\
HOPS-323 & Binary & 273.6 & I & A+B & HC & 123 & 224 & 79 \\
HOPS-366 & Binary & 61.1 & I & A+B & HC & 317 & 37 & 80 \\
HOPS-361 & Binary & 37.6 & 0 & C-A+C-B & HC & 21 & 288 & 87 \\
HOPS-364 & Binary & 65.5 & I & A+B & HC & 96 & 301 & 24 \\
HH270VLA1 & Binary & 99.4 & 0 & A+B & HC & 230 & 124 & 74 \\
\enddata
\tablecomments{\tablenotemark{a}Whether the outflow is launched from a binary or tertiary source. For tertiaries we do not calculate Binary PA or $\Delta$PA. \tablenotemark{b}For tertiaries this is the separation from the inner binary. \tablenotemark{c}These data are from Table 3 and Table 4 in \cite{Tobin_et_al_2022}. \tablenotemark{d}If two sources are listed with a plus we could not distinguish from which source the outflow is driven. Letters connected by a hyphen (e.g., A‑A, A‑B) designate individual components of hierarchical systems. \tablenotemark{e}For all outflow angle measurements, we adopt an uncertainty of 10$^\circ$ for outflows with HC and an uncertainty of 30$^\circ$ for outflows with LC. \tablenotemark{f}This is the average of the red and blue lobes.}
\label{table:by_ouflow}
\end{deluxetable*}

\section{Discussion} \label{sec:dis}
Close-companion protostars are thought to form either directly at $\lesssim$ 500 au scales via disk fragmentation, or they form on $>$1000 au scales via turbulent fragmentation and then migrate to separations $\lesssim$500 au. 

In disk fragmentation, the initial free-fall collapse proceeds straightforwardly to produce a protostellar disk. The disk then fragments via gravitational instability to form a companion, and less commonly it can fragment into multiple companions  \citep{Tobin2016nature, Bate_2018}. Disk fragmentation is expected to preferentially produce close-companion protostellar systems on disk scales ($\lesssim$500 au) with initially aligned circum-multiple disks and outflows ($\Delta$PA $\sim$90$^\circ$) since they have formed from a common structure \citep{Bate_2018}.

Turbulent fragmentation occurs during core collapse, where turbulence produces over-densities that independently collapse. Turbulent fragmentation will preferentially produce wide-separation protostellar systems on core scales ($>$1000 au) with uncorrelated angular momenta due to the stochastic nature of the turbulent flow \citep{Offner_et_al_2016, LeeOffner2019}. Numerical simulations suggest that inward migration of YSOs from separations $>$1000 au to $\lesssim$500 au can occur on timescales of order 10-100 kyr \citep{Offner_et_al_2010, Bate_2012, LeeOffner2019, Kuruwita&Haugbolle2023}. This migration timescale is comparable to the combined protostellar Class 0 and I lifetime of $\sim$500 kyr \citep{Dunham_2014}, meaning that migration may have already taken place in some of our observed systems. Therefore, some of our close-companions may have formed from disk fragmentation and others migrated inward after turbulent fragmentation, and the outflow alignment may be a useful discriminant.

\subsection{Alignment Interpretations}
Figure \ref{fig:histogram} shows a clear preference for values of $\Delta$PA near 90$^\circ$ in our sample, and Figure \ref{fig:cumulat} suggests that a 94\% orthogonal model ($\Delta$PA = 90$^\circ$) would explain the data. These results would favor disk fragmentation over turbulent fragmentation with migration when neglecting any reorientation of the outflows as they migrate inward. However, in systems formed from turbulent fragmentation and undergoing migration, subsequent accretion from a reformed circumbinary disk could torque and similarly bring the outflows into alignment, erasing evidence of the earlier misaligned configuration \citep{Guzejnov2023}. In this scenario, systems formed from turbulent fragmentation and migration could now exhibit well-ordered, near-orthogonal outflows as observed in our sample. This complicates the simple interpretation that the dominantly orthogonal distribution of $\Delta$PA is a direct result of disk fragmentation. Indeed, numerical simulations have reproduced the observed bimodality in separation using only turbulent fragmentation and migration \citep{LeeOffner2019, Kuruwita&Haugbolle2023}.

If disk fragmentation is truly the dominant ($>$71\%) pathway for forming close-companions, we may expect to observe many fragmenting disks among young systems. Yet such cases appear observationally uncommon, with Per-emb-33 (L1448 IRS3B) remaining the best example \citep{Tobin2016nature}. On the other hand, the disk surveys are shallow observations that may miss remnant circum-binary/multiple material due to lower surface brightness sensitivity.

In addition, the broader population of protostellar disks appears largely stable, suggesting that fragmentation may occur only briefly and thus be difficult to observe directly \citep{Tobin_2020}. Moreover, the typical dust disk radii for Class 0 and I YSOs in Orion are $\sim$45 au and $\sim$37 au, respectively, values consistent with those in Perseus \citep{Tobin_2020}. These compact disks might seem inconsistent with producing companions out to $\sim$500 au separations. However, such observations are also compatible with large, gravitationally unstable disks that form and fragment rapidly and early in the star formation process.

\subsection{Close-Companion Binaries with One Outflow} \label{sec:dis:binaries-one-outflow}
Nearly all of our measured sample, 90\% (34/38), contain a close binary system that has only one observable outflow. This number includes the inner binaries in the five hierarchical triple systems. In four of these 34 close binaries, the outflow appears to be launched exclusively by one source (HOPS-12, HOPS-32, HOPS-395, HOPS-400). In the other 30 close binaries, we do not know the extent to which each source is contributing to the outflow; the outflow either appears to launch from directly between the two protostars, or the two protostars are too close to resolve which one is launching the outflow. 

One possibility is that these systems exhibit collective outflows. A numerical simulation in \cite{Peters_et_al_2014} produces collective outflows with poor collimation as a result of variations in angular momentum among the protostars launching the outflows; however, this simulation is high mass turbulent fragmentation and does not fully translate to our sample. We do see evidence of collective outflows in six systems in the form of overlapping aligned outflows. In these instances, there are two distinct overlapping v-shapes pointing in the same direction, often one clearly coming from each source. These systems are Per-emb-22, HOPS-92, HOPS-84, HOPS-182, HOPS-312, and HOPS-323. In these systems, the collective outflows may be less collimated, making the overlapping structures more apparent, whereas other systems with more tightly collimated collective outflows remain unresolved. A collective outflow would indicate that the two protostars have aligned angular momenta, consistent with forming from the same disk. Additionally, there is some empirical evidence showing that close binaries with a single outflow detected in molecular lines actually have two parallel jets that are visible with different tracer \citep{Tychoniec+2024, Barsony+2024}. However, we cannot conclude here that only detecting one outflow indicates that the other outflow is aligned and superimposed.

% Classifications
Following the results of our modeling (Figure \ref{fig:cumulat}), which is based solely on outflow PA and binary separation PA alignments, these close binary systems launching a single outflow are consistent with being dominantly formed from disk fragmentation, and the presence of collective outflows may provide further support for disk fragmentation. \cite{Reynolds_et_al_2021, Reynolds_et_al_2024} classified five of the 34 close binaries in this section as disk fragmentation candidates. Their classifications are also based on ALMA Band 6 (1.3 mm) data. Per-emb-2, Per-emb-17, Per-emb-18, Per-emb-22, and Per-emb-33 were classified as likely disk fragmentation candidates due to strong disk alignment and bright circum-multiple material in some cases. Our findings are consistent with these classifications. However, we cannot conclusively attribute all 34 close binaries to disk fragmentation, as alternative formation scenarios (such as those discussed above) may produce similar alignment signatures.

\subsection{Close-Companion Binaries with Two Outflows} \label{sec:dis:binaries-two-outflows}
Four of 38 of our measured samples are close binary systems where each companion launches its own distinct outflow. These are Per-emb-35, Per-emb-27, Per-emb-12, and HOPS-290. These cases provide us with another way to infer the formation mechanism: the alignment between the outflow PAs of both protostars.

% Classifications
In the case of Per-emb-12, the red lobes seem to be merging collectively while the blue lobes appear to be interacting with each other \citep[e.g.,][]{Encalada_2024}. The small misalignment (28$^\circ$) of the two outflow PAs and the nuances of the morphology suggest that we can consider the outflows of these companions to be aligned and consistent with disk fragmentation. 

The orthogonal outflows in Per-emb-27 (91$^\circ$ apart) and the outflows oriented antiparallel to each other in Per-emb-35 (169$^\circ$ apart) suggest that they were not formed from the same disk, and that turbulent fragmentation and inward migration are more likely. \cite{Tobin_et_al_2016} also notes that the close separation and orthogonal outflows of Per-emb-27 suggest turbulent fragmentation with migration. \cite{Reynolds_et_al_2024} classifies Per-emb-27 and Per-emb-35 as likely turbulent fragmentation candidates because Per-emb-27 has strongly misaligned disks, and Per-emb-35 has a wider separation (570 au) and lacks circum-multiple material. 

The outflows in HOPS-290 are misaligned (65$^\circ$ apart). Additionally, in the dust continuum, HOPS-290-B is elongated while HOPS-290-A is not, suggesting the disks are also misaligned. This is more consistent with turbulent fragmentation and inward migration.

\subsection{Hierarchical Triples with Two Outflows} \label{sec:dis:hierarchical-triples}
Five of the 38 in our measured sample are hierarchical triple systems where a tertiary source launches its own outflow. These systems are Per-emb-33, HOPS-12, HOPS-92, HOPS-203, and HOPS-288. The differences between the PAs of the inner binary outflow and the tertiary outflow are 15$^\circ$, 0$^\circ$, 167$^\circ$, 43$^\circ$, and 156$^\circ$, respectively. 

% Classifications
HOPS-12 is a peculiar case because its tertiary source lies $\sim$1800 au from the inner binary, yet the outflow PAs are aligned within 1$^\circ$. However, outflow alignment on larger scales like this is not extraordinary. For example, three outflows in the larger L1448 region (Per-emb-33, Per-emb-22, and L1448 IRS3C) are aligned within $\sim$6$^\circ$, and 20 outflows in Serpens Main are aligned within 24$^\circ$ \citep{Green_2024}. It is also possible that these wide binaries are aligned because they have inherited the angular momentum of their core \citep{Sadavoy_Stahler_2017}. This tertiary companion is a turbulent fragmentation candidate since disk fragmentation is not possible on large scales.

The outflow PAs of the tertiary companions in HOPS-92 and HOPS-288 are both misaligned to the outflow PA of the inner binary, but the tertiaries are close-companions to the inner binary. These may be turbulent fragmentation with inward migration candidates. 

The outflow PA misalignment of the HOPS-203 tertiary companion is ambiguous (43$^\circ$) and the separation from the inner binary is not close, at $\sim$1000 au. This tertiary companion is a turbulent fragmentation candidate since disk fragmentation is not possible on large scales.

The outflows on the tertiary and the inner binary of Per-emb-33 (L1448 IRS3B) are aligned within 15$^\circ$. The dust continuum shows three sources with circum-multiple material in Keplerian rotation, and a marginally unstable disk \citep{Tobin2016nature, Reynolds_et_al_2021}. Per-emb-33 is a prime example of disk fragmentation.

\section{Conclusions} \label{sec:conc}
From our large sample of 51 close-companion Class 0/I protostellar systems, based on our measurements of 42 outflow PAs in 38 systems, and the preceding discussion, we draw attention to the following conclusions:

\begin{itemize}
    \item In 38 of 51 systems ($\sim$75\%) we see identifiable and measurable molecular outflows. In 34 of the 38 systems ($\sim$89\%) we see a close binary that drives one outflow. We can not distinguish if the outflow is from one or both protostars, but the presence of aligned and overlapping outflows driven by distinct sources in some systems suggests a collective origin. Additionally, in five of these 34 systems, we see tertiary companions that launch their own outflow, in addition to the outflow launched by the inner binary. In the remaining four of 38 systems ($\sim$25\%), we measure two outflows where each companion drives its own outflow. 
    \item KS tests show that our observed $\Delta$PA (difference between the outflow PA and binary separation PA) distribution is consistent with being at least $\sim$71\% orthogonally aligned at a significance level of 0.05. We interpret this to mean that most binaries exhibit outflows launching at 90$^\circ$ to the binary orbital plane, which could be evidence of disk fragmentation or rapid alignment of merging systems.
    \item We posit that disk fragmentation is likely the dominant formation pathway for close-companion protostellar systems, and turbulent fragmentation followed by inward migration is a less common formation pathway. However, the scarcity of observed fragmenting disks and the compact nature of most protostellar disks may suggest more complication in this inference.
\end{itemize}

\clearpage
%% IMPORTANT! The old "\acknowledgment" command has be depreciated. It was
%% not robust enough to handle our new dual anonymous review requirements and
%% thus been replaced with the acknowledgment environment. If you try to 
%% compile with \acknowledgment you will get an error print to the screen
%% and in the compiled pdf.
%% 
%% Also note that the akcnowlodgment environment does not support long amounts of text. If you have a lot of people and institutions to acknowledge, do not use this command. Instead, create a new \section{Acknowledgments}.
\nolinenumbers
\begin{acknowledgments}
This paper makes use of the following ALMA data: ADS/JAO.ALMA$\#$2018.1.01038.S and ADS/JAO.ALMA$\#$2017.1.00053.S. ALMA is a partnership of ESO (representing its member states), NSF (USA) and NINS (Japan), together with NRC (Canada), NSTC and ASIAA (Taiwan), and KASI (Republic of Korea), in cooperation with the Republic of Chile. The Joint ALMA Observatory is operated by ESO, AUI/NRAO and NAOJ. The National Radio Astronomy Observatory and Green Bank Observatory are facilities of the U.S. National Science Foundation operated under cooperative agreement by Associated Universities, Inc. L.W.L. and R.S. acknowledges support from NSF AST 2108794. F.J.E. acknowledges support from NRAO SOSPA6-010 363369. J.J.T. acknowledges support from NASA XRP 80NSSC22K1159. N.M.M. acknowledges support from  DGAPA–PAPIIT IA103025. 

This research made use of APLpy, an open-source plotting package for Python \citep{aplpy2012, aplpy2019}. This work made use of Astropy, a community-developed core Python package and an ecosystem of tools and resources for astronomy \citep{astropy:2013, astropy:2018, astropy:2022}.

The figures, tables, scripts used to make them, and FITS images are available in the Illinois Data Bank \citep{illinoisdatabankIDB-7093951}. The ALMA data and reduction scripts are available from the ALMA archive.
\end{acknowledgments}

%% To help institutions obtain information on the effectiveness of their 
%% telescopes the AAS Journals has created a group of keywords for telescope 
%% facilities.
%
%% Following the acknowledgments section, use the following syntax and the
%% \facility{} or \facilities{} macros to list the keywords of facilities used 
%% in the research for the paper.  Each keyword is check against the master 
%% list during copy editing.  Individual instruments can be provided in 
%% parentheses, after the keyword, but they are not verified.

\vspace{5mm}
\facilities{ALMA}

%% Similar to \facility{}, there is the optional \software command to allow 
%% authors a place to specify which programs were used during the creation of 
%% the manuscript. Authors should list each code and include either a
%% citation or url to the code inside ()s when available.

\software{Astropy \citep{astropy:2013, astropy:2018, astropy:2022}, APLpy \citep{aplpy2012, aplpy2019}, CARTA \citep{carta}, CASA \citep{CASA}} 

%% Appendix material should be preceded with a single \appendix command.
%% There should be a \section command for each appendix. Mark appendix
%% subsections with the same markup you use in the main body of the paper.

%% Each Appendix (indicated with \section) will be lettered A, B, C, etc.
%% The equation counter will reset when it encounters the \appendix
%% command and will number appendix equations (A1), (A2), etc. The
%% Figure and Table counter will not reset.

%% For this sample we use BibTeX plus aasjournals.bst to generate the
%% the bibliography. The sample631.bib file was populated from ADS. To
%% get the citations to show in the compiled file do the following:
%%
%% pdflatex sample631.tex
%% bibtext sample631
%% pdflatex sample631.tex
%% pdflatex sample631.tex

\bibliography{main}{}
\bibliographystyle{aasjournal}

\appendix

\section{All Fields with Notes} \label{sec:appendix}

\subsection{Overview and Descriptive Terminology} \label{sec:appendix:overview}
We present the $^{12}$CO ($J=2\rightarrow1$) maps of all 51 fields in two sections: fields with outflow emission and fields without outflow emission. Each section lists the fields in order of increasing Right Ascension. Fields with at least one measured outflow are displayed in integrated intensity maps (moment 0) in which we selectively integrated channels that highlight the outflow morphology. Channels near the systemic velocity were omitted to better show the outflow material. Data below 3 ${\sigma}$ were clipped to guarantee the significance of the data. These fields are annotated with the binary separation PA and the outflow PA(s), and markers indicating the source position. Some source markers may be slightly offset due to proper motion or since the original coordinates are from VLA 9~mm data which might have a slightly different centroid position than these ALMA 1.3~mm data. The moment maps were created with self-written python scripts. Table \ref{table:by_field} notes which velocity ranges were integrated. Fields with no measured outflows are displayed with maximum intensity maps (moment 8) and no clipping. The maps are provided as a figure set at the end of the appendix.

In this Appendix, each field has a paragraph describing the characteristics of the system, the outflow morphology, and any other nuances or idiosyncrasies. We call an outflow sharp if its edges are clearly defined and contained, and diffuse if its edges are not clearly defined or blurry. In some cases, we note that there may be an overlapping aligned outflow present. This means that there is evidence of two overlapping v-shapes pointing in the same direction. This may be hinting at the idea of collective outflows. However, in each of these cases, we treat it as a single outflow. Please refer to \ref{sec:results:methods} for additional definitions of terms used in the qualitative descriptions in this appendix. Outflows that we attribute as LC are explicitly noted and reasons provided. Every other outflow is HC. Some of the characteristics described may not be visible in the maps provided since they were constructed with velocity channels that best reveal the outflow. Outflows in systems with more than one outflow may appear tangled up in the moment maps but they have been disentangled and measured separately by looking at specific velocity ranges. Additionally, the Perseus outflows note measurements from prior studies for comparison.

\begin{deluxetable*}{llllll}
\tablenum{2}
\tablecaption{Data by Field}
\tablehead{
    \colhead{Field} & 
    \colhead{Other Name} & 
    \colhead{\# of Sources} & 
    \colhead{RA\tablenotemark{a}\ (ICRS)} & 
    \colhead{Dec\tablenotemark{a}\ (ICRS)} & 
    \colhead{Integrated Velocities\tablenotemark{b}\ (km~s$^{-1}$)}
}
\startdata
Per-emb-22 & L1448 IRS2 & 2 & 51.34337 & 30.75368 & 7.3 to 17.2, -3.92 to 0.04 \\
L1448 IRS3C & \nodata & 2 & 51.39862 & 30.75948 & 6.26 to 14.84, -1.0 to 2.96 \\
Per-emb-33 & L1448 IRS3B & 3 & 51.40159 & 30.75409 & 6.64 to 13.24 \\
Per-emb-17 & \nodata & 2 & 51.91293 & 30.21752 & 7.96 to 11.92, 0.04 to 2.68 \\
Per-emb-35 & \nodata & 2 & 52.15454 & 31.22522 & 9.94 to 12.58 \\
Per-emb-27 & NGC 1333 IRS2A & 2 & 52.23154 & 31.24362 & 11.26 to 14.56, -1.28 to 3.34 \\
Per-emb-36 & NGC 1333 IRAS2B & 2 & 52.23906 & 31.23771 & 11.92 to 15.22, 0.04 to 3.34 \\
Per-emb-44 & NGC 1333 SVS 13A & 2 & 52.26569 & 31.26772 & 16.54 to 20.5, -33.62 to -26.36 \\
Per-emb-12 & NGC 1333 IRAS4A & 2 & 52.2939 & 31.22526 & 9.94 to 13.24, -2.6 to 2.02 \\
Per-emb-18 & \nodata & 2 & 52.29691 & 31.30863 & 10.6 to 13.9 \\
Per-emb-2 & IRAS 03292+3039 & 5 & 53.07471 & 30.82992 & 9.28 to 12.58, 1.36 to 4.0 \\
HOPS-32 & \nodata & 2 & 83.64769 & -5.66653 & 10.0 to 14.0, 3.0 to 6.0 \\
HOPS-28 & \nodata & 2 & 83.69714 & -5.69895 & \nodata \\
HOPS-45 & \nodata & 2 & 83.77691 & -5.55976 & \nodata \\
HOPS-12 & \nodata & 3 & 83.7873 & -5.93194 & 11.0 to 15.0, 0.0 to 5.0 \\
HOPS-92 & \nodata & 3 & 83.82641 & -5.00916 & 14.0 to 20.0, 1.0 to 5.0 \\
HOPS-56 & \nodata & 4 & 83.83119 & -5.25916 & \nodata \\
HOPS-70 & \nodata & 5 & 83.84341 & -5.13476 & \nodata \\
HOPS-84 & \nodata & 2 & 83.86067 & -5.06532 & 12.0 to 15.0, 6.0 to 8.0 \\
HOPS-75 & \nodata & 2 & 83.86118 & -5.10297 & 14.0 to 16.0 \\
HOPS-85 & \nodata & 2 & 83.86747 & -5.06147 & \nodata \\
HOPS-77 & \nodata & 3 & 83.88143 & -5.0965 & \nodata \\
HOPS-182 & \nodata & 3 & 84.07828 & -6.36963 & 12.0 to 15.0, -4.0 to 1.0 \\
HOPS-168 & \nodata & 2 & 84.07896 & -6.75655 & 12.0 to 23.0, -1.0 to 6.0 \\
HOPS-203 & \nodata & 3 & 84.09528 & -6.76852 & 4.0 to 8.0 \\
HOPS-173 & \nodata & 2 & 84.10853 & -6.41813 & 10.0 to 10.0 \\
HOPS-193 & \nodata & 2 & 84.12619 & -6.02146 & 12.0 to 12.0, 5.0 to 7.0 \\
HOPS-163 & \nodata & 2 & 84.32199 & -6.60506 & \nodata \\
HOPS-158 & \nodata & 2 & 84.352 & -6.97582 & 9.0 to 11.0, -2.0 to 3.0 \\
HOPS-138 & \nodata & 2 & 84.70138 & -7.04548 & \nodata \\
HOPS-395 & \nodata & 2 & 84.82122 & -7.40683 & 6.0 to 6.0, 2.0 to 3.0 \\
HOPS-288 & \nodata & 3 & 84.98333 & -7.50766 & 10.0 to 16.0, -7.0 to -1.0 \\
HOPS-290 & \nodata & 2 & 84.98888 & -7.49242 & 7.0 to 10.0, -1.0 to 2.0 \\
HOPS-281 & \nodata & 2 & 85.10262 & -7.71897 & 7.0 to 13.0, -1.0 to 2.0 \\
HOPS-282 & \nodata & 2 & 85.1087 & -7.62559 & 7.0 to 9.0, 1.0 to 3.0 \\
HOPS-242 & \nodata & 2 & 85.20226 & -8.18582 & \nodata \\
HOPS-255 & \nodata & 2 & 85.21074 & -8.09687 & \nodata \\
HOPS-261 & \nodata & 2 & 85.32872 & -7.92471 & \nodata \\
HOPS-248 & \nodata & 2 & 85.34223 & -7.96752 & \nodata \\
HOPS-357 & \nodata & 2 & 85.41307 & -1.86872 & \nodata \\
HOPS-384 & \nodata & 10 & 85.43391 & -1.9128 & 16.0 to 20.0 \\
HOPS-304 & \nodata & 2 & 85.44139 & -1.94076 & 15.0 to 17.0, 2.0 to 6.0 \\
HOPS-400 & \nodata & 2 & 85.68856 & -1.27052 & 11.0 to 16.0, 3.0 to 6.0 \\
HOPS-213 & \nodata & 2 & 85.70035 & -8.669 & 5.0 to 6.0, 0.0 to 1.0 \\
HOPS-312 & \nodata & 2 & 85.77375 & -1.26521 & 5.0 to 6.0, -4.0 to -1.0 \\
HOPS-363 & \nodata & 2 & 86.67969 & 0.01453 & 15.0 to 17.0, 0.0 to 1.0 \\
HOPS-323 & \nodata & 2 & 86.69863 & 0.00689 & 13.0 to 15.0, 2.0 to 7.0 \\
HOPS-366 & \nodata & 2 & 86.7667 & 0.36955 & 20.0 to 28.0, -12.0 to -5.0 \\
HOPS-361 & \nodata & 13 & 86.76931 & 0.36328 & -11.0 to -4.0 \\
HOPS-364 & \nodata & 2 & 86.90235 & 0.33499 & 14.0 to 17.0, 3.0 to 7.0 \\
HH270VLA1 & \nodata & 2 & 87.89415 & 2.94608 & 12.0 to 17.0, 2.0 to 8.0 \\
\enddata
\tablecomments{\tablenotemark{a}These are the field center coordinates used in the maps. \tablenotemark{b}Fields with no integrated velocities are fields in which no outflows were measured; we display them with maximum intensity (moment 8) maps.}
\label{table:by_field}
\end{deluxetable*}

\begin{deluxetable*}{lllllll}
\tablenum{3}
\tablecaption{Outflow PAs}
\tablehead{
    \colhead{Field} & 
    \colhead{Source} & 
    \colhead{Conf.\tablenotemark{a}} & 
    \colhead{Red Lobe PA\tablenotemark{b}\ ($^\circ$)} & 
    \colhead{Blue Lobe PA\ ($^\circ$)} & 
    \colhead{Lobe Offset\tablenotemark{c}\ ($^\circ$)} & 
    \colhead{Average PA\ ($^\circ$)}
}
\startdata
Per-emb-22 & A+B & HC & 313 & 305 & 8 & 309 \\
L1448 IRS3C & A+B & HC & 310 & 304 & 6 & 307 \\
Per-emb-33 & A+B & HC & 303 & \nodata & \nodata & 303 \\
Per-emb-33 & C & HC & 288 & \nodata & \nodata & 288 \\
Per-emb-17 & A+B & HC & 228 & 244 & 16 & 236 \\
Per-emb-35 & A & HC & \nodata & 119 & \nodata & 119 \\
Per-emb-35 & B & HC & 313 & 308 & 5 & 310 \\
Per-emb-27 & A & HC & 191 & 215 & 24 & 203 \\
Per-emb-27 & B & HC & 294 & \nodata & \nodata & 294 \\
Per-emb-36 & A+B & LC & 244 & 208 & 36 & 226 \\
Per-emb-44 & A+B & LC & 163 & 115 & 48 & 139 \\
Per-emb-12 & B & HC & 205 & 211 & 6 & 208 \\
Per-emb-12 & A & HC & 190 & 171 & 19 & 180 \\
Per-emb-18 & A+B & HC & 167 & \nodata & \nodata & 167 \\
Per-emb-2 & B & HC & 127 & 125 & 2 & 126 \\
HOPS-32 & B & HC & 341 & 337 & 4 & 339 \\
HOPS-12 & A & HC & \nodata & 337 & \nodata & 337 \\
HOPS-12 & B-A & HC & 336 & 338 & 2 & 337 \\
HOPS-92 & B & HC & \nodata & 89 & \nodata & 89 \\
HOPS-92 & A-A+A-B & HC & 256 & \nodata & \nodata & 256 \\
HOPS-84 & A+B & HC & 261 & 264 & 3 & 263 \\
HOPS-75 & A+B & LC & 280 & \nodata & \nodata & 280 \\
HOPS-182 & A+B & HC & 47 & 63 & 16 & 55 \\
HOPS-168 & A+B & HC & 337 & 342 & 5 & 339 \\
HOPS-203 & A+B & HC & \nodata & 324 & \nodata & 324 \\
HOPS-203 & C & HC & \nodata & 281 & \nodata & 281 \\
HOPS-173 & A+B & LC & 67 & \nodata & \nodata & 67 \\
HOPS-193 & A+B & HC & 278 & 269 & 8 & 274 \\
HOPS-158 & A+B & LC & 47 & 65 & 18 & 56 \\
HOPS-395 & B & HC & 347 & 1 & 14 & 174 \\
HOPS-288 & A-A+A-B & HC & \nodata & 236 & \nodata & 236 \\
HOPS-288 & B & HC & 32 & \nodata & \nodata & 32 \\
HOPS-290 & A & HC & 116 & 112 & 4 & 114 \\
HOPS-290 & B & HC & 50 & 48 & 3 & 49 \\
HOPS-281 & A+B & HC & 207 & 205 & 3 & 206 \\
HOPS-282 & A+B & HC & 1 & 5 & 4 & 3 \\
HOPS-384 & A+A-B & LC & 4 & \nodata & \nodata & 4 \\
HOPS-304 & A+B & LC & 167 & 136 & 30 & 152 \\
HOPS-400 & A & HC & 77 & 96 & 19 & 87 \\
HOPS-213 & A+B & LC & 87 & 82 & 5 & 84 \\
HOPS-312 & A+B & HC & 36 & 37 & 1 & 36 \\
HOPS-363 & A+B & LC & 285 & 261 & 24 & 273 \\
HOPS-323 & A+B & HC & 127 & 119 & 7 & 123 \\
HOPS-366 & A+B & HC & \nodata & 317 & \nodata & 317 \\
HOPS-361 & C-A+C-B & HC & \nodata & 21 & \nodata & 21 \\
HOPS-364 & A+B & HC & 88 & 105 & 18 & 96 \\
HH270VLA1 & A+B & HC & 230 & 230 & 1 & 230 \\
\enddata
\tablecomments{\tablenotemark{a}For all outflow angle measurements, we adopt an uncertainty of 10$^\circ$ for outflows with HC and an uncertainty of 30$^\circ$ for outflows with LC. \tablenotemark{b}The red lobe PA is given flipped by 180$^\circ$ for easy comparison with the blue lobe PA. \tablenotemark{c}The difference between red and blue lobe PAs, if applicable.}
\label{table:by_outflow_full}
\end{deluxetable*}

\begin{deluxetable*}{lllllll}
\tablenum{4}
\tablecaption{Perseus Source Position Offsets}
\tablehead{
    \colhead{Field} & 
    \colhead{Source} & 
    \colhead{RA Offset\ (arcsec)} &
    \colhead{Dec Offset\ (arcsec)}
}
\startdata
Per-emb-22 & Per-emb-22-B & 0.08 & -0.025 \\
Per-emb-22 & Per-emb-22-A & 0.18 & -0.025 \\
L1448 IRS3C & L1448 IRS3C-A & -0.05 & 0.05 \\
L1448 IRS3C & L1448 IRS3C-B & -0.05 & 0.05 \\
Per-emb-33 & Per-emb-33-B & 0.1 & 0.0 \\
Per-emb-33 & Per-emb-33-A & 0.0 & 0.0 \\
Per-emb-33 & Per-emb-33-C & 0.05 & 0.05 \\
Per-emb-17 & Per-emb-17-A & 0.125 & -0.075 \\
Per-emb-17 & Per-emb-17-B & 0.125 & -0.075 \\
Per-emb-35 & Per-emb-35-A & 0.05 & 0.0 \\
Per-emb-35 & Per-emb-35-B & 0.0 & 0.0 \\
Per-emb-27 & Per-emb-27-B & 0.05 & -0.05 \\
Per-emb-27 & Per-emb-27-A & 0.0 & 0.0 \\
Per-emb-36 & Per-emb-36-B & 0.0 & -0.1 \\
Per-emb-36 & Per-emb-36-A & 0.1 & -0.075 \\
Per-emb-44 & Per-emb-44-B & 0.075 & -0.05 \\
Per-emb-44 & Per-emb-44-A & 0.1 & -0.05 \\
Per-emb-12 & Per-emb-12-B & 0.0 & -0.1 \\
Per-emb-12 & Per-emb-12-A & 0.1 & -0.1 \\
Per-emb-18 & Per-emb-18-B & 0.0 & 0.05 \\
Per-emb-18 & Per-emb-18-A & 0.2 & 0.1 \\
Per-emb-2 & Per-emb-2-E & 0.0 & 0.0 \\
Per-emb-2 & Per-emb-2-B & 0.0 & 0.0 \\
Per-emb-2 & Per-emb-2-C & 0.0 & 0.0 \\
Per-emb-2 & Per-emb-2-A & 0.0 & 0.0 \\
Per-emb-2 & Per-emb-2-D & 0.0 & 0.0 \\
\enddata
\tablecomments{The Perseus source positions are offset slightly from the reported positions in \cite{Tobin_et_al_2018}. The offsets in this table are corrections to those source positions and were applied in our figures.}
\label{table:offsets}
\end{deluxetable*}

\subsection{Fields with Outflow Emission} \label{sec:appendix:fields-with-emission}

\subsubsection{Per-emb-22}
Per-emb-22 (also known as L1448 IRS2) contains two sources with a separation of 225 au. We measured one HC outflow with two lobes. The lobe-averaged PA is 129$^\circ$, the lobe PA offset is 8$^\circ$, and the launching source is not distinguishable. The blue lobe is a sharp v-shape that launches from Per-emb-22-B. Per-emb-22-A has much fainter emission that is potentially also contributing to the outflow, or its own overlapping and aligned outflow since the red lobe appears to have sharp aligned v-shapes launching from both sources, but they are ambiguous. Ultimately, due to ambiguity, we treat it as a single outflow and measure one PA.

\cite{Stephens_et_al_2017} measured an outflow PA of 118$^\circ$, which differs from our measurement by 11$^\circ$. \cite{Reynolds_et_al_2024} classify this system as a likely disk fragmentation candidate due to the aligned orientations of the components.

\subsubsection{L1448 IRS3C}
L1448 IRS3C (also known as L1448NW) contains two sources with a separation of 75 au. We measured one HC outflow with two lobes. The lobe-averaged PA is 127$^\circ$, and the lobe PA offset is 6$^\circ$. The blue lobe is a diffuse v-shape that is only visible in a few channels. The red lobe is a diffuse v-shape with higher signal-to-noise than the blue lobe. The outflow launching source is not distinguishable.

\subsubsection{Per-emb-33}
Per-emb-33 (also known as L1448 IRS3B) contains three sources in a hierarchical triple system in which we measure two HC outflows. \cite{Reynolds_et_al_2024}, however, detect a fourth compact continuum source that may be associated with a central component of the system. Also in the field is the wide ($>$ 2000 au) companion L1448 IRS3A, but we do not include it in this analysis. The inner binary (Per-emb-33-A and Per-emb-33-B) has a separation of 80 au and the tertiary (Per-emb-33-C) is 263 au from the inner binary. All three sources are embedded within a spiral structure seen in continuum. In the blue channels, we cannot confidently detect an outflow lobe, although there is an unambiguous jet-like structure launching from the system. In the red channels, both the binary and tertiary appear to launch separate outflows, but they are overlapping. The binary has a broader sharp v-shape, and we cannot distinguish which binary source launches the outflow. The tertiary appears in projection against the outflow cavity and is launching its own sharp v-shaped outflow. This outflow is narrower and lies entirely within the cavity of the binary's outflow. Both outflows share a northern edge. The outflow PA of the binary is 123$^\circ$. The outflow PA of the tertiary is 108$^\circ$. 

\cite{Lee_2015} measured a single outflow PA of 122$^\circ$, corresponding to the wider outflow from the inner binary, which differs from our measurement by 1$^\circ$. \cite{Reynolds_et_al_2021} examine this system in more detail, also noting an outflow, and conclude that disk fragmentation is the most likely formation mechanism.

\subsubsection{Per-emb-17}
Per-emb-17 contains two sources with a separation of 83 au. We measured one HC outflow with two lobes. The lobe-averaged PA is 56$^\circ$, the lobe PA offset is 16$^\circ$, and the launching source is not distinguishable. The blue lobe is a diffuse v-shape with some extraneous emission structure in the cavity. The red lobe is a sharp v-shape.

\cite{Stephens_et_al_2017} measure an outflow PA of 57$^\circ$, which differs from our measurement by 1$^\circ$. \cite{Dunham_et_al_2024} measure an outflow PA of 64$^\circ$, which differs from our measurement by 8$^\circ$. They also determine an outflow opening angle of 69$^\circ$ $\pm$ 8$^\circ$. \cite{Reynolds_et_al_2024} classify this system as a likely disk fragmentation candidate.

\subsubsection{Per-emb-35}
Per-emb-35 contains two sources with a separation of 573 au. We measured two HC outflows. On Per-emb-35-A we observe a diffuse v-shaped blue lobe and do not detect a red lobe. The outflow PA is 119$^\circ$. On Per-emb-35-B, the northwest lobe is visible in both red and blue channels, and we assign this lobe as the blue lobe since the red channels also contain the southeast lobe. The lobe-averaged PA is 310$^\circ$ and the lobe PA offset is 5$^\circ$.

\cite{Stephens_et_al_2017} measure two outflow PAs of 123$^\circ$ and 349$^\circ$. These differ from our measurements by 4$^\circ$ and 39$^\circ$ respectively. \cite{Reynolds_et_al_2024} classify this system as a turbulent fragmentation candidate.

\subsubsection{Per-emb-27}
Per-emb-27 (also known as NGC 1333 IRS2A) contains two sources with a separation of 186 au. We measured two HC outflows. Per-emb-27-A launches an outflow with two sharp v-shaped lobes. The lobe-averaged PA is 23$^\circ$ and the lobe PA offset is 24$^\circ$. Per-emb-27-B launches a second outflow, roughly perpendicular to the outflow of Source A. This outflow is slightly peculiar, in that it has a distorted and narrow shape where the edges appear to twist around each other and alter the pointing direction. We only detect a red lobe for Source B, and the outflow PA is 114$^\circ$.

\cite{Plunkett_2013} measure two outflow PAs of 14$^\circ$ and 104$^\circ$, differing from our measurements by 9$^\circ$ and 10$^\circ$ respectively. \cite{Dunham_et_al_2024} only measure the outflow associated with Per-emb-27-A, at 14$^\circ$, in agreement with \cite{Plunkett_2013} and differing from our measurement by 9$^\circ$. \cite{Dunham_et_al_2024} also determine an outflow opening angle of 65$^\circ$ $\pm$ 11$^\circ$. \cite{Reynolds_et_al_2024} classify this system as a turbulent fragmentation candidate.

\subsubsection{Per-emb-36}
Per-emb-36 (also known as NGC 1333 IRAS2B) contains two sources with a separation of 93 au. We measured one LC outflow with two lobes. The lobe-averaged PA is 46$^\circ$, the lobe PA offset is 36$^\circ$, and the launching source is not distinguishable. The blue lobe is extremely wide at 162$^\circ$ $\pm$ 31$^\circ$ \citep{Dunham_et_al_2024}, and the red lobe is a diffuse v-shape. There is extraneous emission north of the system that does not appear to be related to an outflow. Due to the wide outflow and large lobe PA offset, we label our measurements LC.

\cite{Plunkett_2013} measure an outflow PA of 24$^\circ$, and \cite{Dunham_et_al_2024} measure an outflow PA of 13$^\circ$ and the outflow opening angle of 162$^\circ$ $\pm$ 31$^\circ$. These differ from our measurement by 22$^\circ$ and 33$^\circ$ respectively. \cite{Reynolds_et_al_2024} classify the formation mechanism of this system as inconclusive.

\subsubsection{Per-emb-44}
Per-emb-44 (also known as NGC 1333 SVS 13A) contains two sources with a separation of 90 au. We measured one LC outflow with two lobes. The lobe-averaged PA is 319$^\circ$, the lobe offset PA is 48$^\circ$, and the launching source is not distinguishable. The continuum data exhibits a spiral structure around the binary. The blue channels contain emission that could be attributed to an outflow in velocity channels as high as 69 km~s$^{-1}$. The blue lobe is a wide u-shape, the base of which lies slightly behind the binary sources. The red lobe is a sharp u-shape. Due to the large lobe offset PA, we label these measurements LC.

\cite{Plunkett_2013} measure an outflow PA of 130$^\circ$, as does \cite{Lee_et_al_2016}. These differ from our measurement by 11$^\circ$. \cite{Dunham_et_al_2024} measure an outflow PA of 140$^\circ$, differing from our measurement by 21$^\circ$. They also determine an outflow opening angle of 144$^\circ$ $\pm$ 20$^\circ$. 

\subsubsection{Per-emb-12}
Per-emb-12 (also known as NGC 1333 IRAS4A) contains two sources with a separation of 549 au. We measured two HC outflows. Both Per-emb-12-A and Per-emb-12-B appear to launch outflows, which have sharp v-shapes. The two blue lobes are not overlapping, but appear to be interacting and deflecting in slightly different directions. The two red lobes appear to be overlapping and pointing in the same direction, however they are seen individually launching an outflow near the sources. The lobe-averaged PA of the outflow launched by Per-emb-12-A is 0$^\circ$ and the lobe PA offset is 19$^\circ$. The lobe-averaged PA of the outflow launched by Per-emb-12-B is 28$^\circ$ and the lobe PA offset is 6$^\circ$.

\cite{Plunkett_2013} measure a single outflow PA of 35$^\circ$ and \cite{Lee_et_al_2016} measure a single outflow PA of 19$^\circ$. They attributed only a single outflow PA to this system, whereas we measure two separate outflow PAs, one driven from each source. These compare to our PAs of 0$^\circ$ and 28$^\circ$, which if you averaged as a single outflow would be 14$^\circ$. \cite{Chahine_et_al_2024} delve further into the outflow morphology and kinematics, identifying two outflow systems in each protostar. Their more nuanced analysis incorporated additional tracers and emphasizes a more complex picture of the outflows.

\subsubsection{Per-emb-18}
Per-emb-18 contains two sources with a separation of 26 au. We measured one HC outflow with one lobe. The outflow PA is 347$^\circ$, and the launching source is not distinguishable. We cannot confidently detect a blue lobe, although the blue channels contain some filamentary structure, but this emission is neither in the position where an outflow should be nor resembles an outflow. The red lobe has a sharp v-shape on the north side and also exhibits a diffuse v-shape on the south side of the binary, indicating an edge-on disk. There is extraneous emission with an arc-like structure southwest of the system in the red channels.

\cite{Lee_et_al_2016} measure an outflow PA of 150$^\circ$, differing from our measurement by 17$^\circ$. \cite{Reynolds_et_al_2024} classify this system as a likely disk fragmentation candidate.

\subsubsection{Per-emb-2}
Per-emb-2 contains five sources, of which two form a close-companion system with a separation of 24 au. We measured one HC outflow with two lobes. The lobe-average PA is 306$^\circ$, the lobe PA offset is 2$^\circ$, and the launching source is not distinguishable. Both lobes have a sharp v-shape. In continuum, there is a large cloud-like structure containing four different embedded sources.

\cite{Stephens_et_al_2017} measure an outflow PA of 129$^\circ$, and \cite{Dunham_et_al_2024} measure an outflow PA of 131$^\circ$ along with an outflow opening angle of 40$^\circ$ $\pm$ 8$^\circ$. These differ from our measurements by 5$^\circ$ and 3$^\circ$ respectively. \cite{Reynolds_et_al_2024} classify this system as a likely disk fragmentation candidate.

\subsubsection{HOPS-32}
HOPS-32 contains two sources with a separation of 162 au. We measured one HC outflow with two lobes. The lobe-average PA is 159$^\circ$ and the lobe PA offset is 4$^\circ$. The outflow is launched from HOPS-32-B. The blue lobe exhibits sharp v-shapes on both sides of the binary, indicating an edge-on disk. The red lobe contains a faint and sharp v-shape on the south side of the binary.

\subsubsection{HOPS-45}
HOPS-45 contains two sources with a separation of 121 au. We did not measure any outflows in this field. The blue channels have some sort of compact ring that arcs around the binary with no clear directionality. The red channels contain a possible small v-shaped lobe, but we cannot confidently detect an outflow lobe.

\subsubsection{HOPS-12}
HOPS-12 contains three sources in a hierarchical triple system in which we measure two HC outflows. The inner binary (HOPS-12-B-A and HOPS-12-B-B) has a separation of 81 au and the tertiary (HOPS-12-A) is 1854 au from the inner binary. The inner binary has two sharp v-shaped lobes. The lobe-averaged PA is 157$^\circ$ and the lobe PA offset is 2$^\circ$, and the launching source is not distinguishable. The tertiary has a diffuse u-shaped miniature blue lobe and we cannot confidently detect a red lobe, although there is some ambiguous emission structure. This outflow is very small compared to any other in this sample. The outflow PA is 157$^\circ$, aligned with the outflow of the inner binary.

\subsubsection{HOPS-92}
HOPS-92 contains three sources in a hierarchical triple system in which we measure two HC outflows, each only having one lobe. The inner binary (HOPS-92-A-A and HOPS-92-A-B) has a separation of 108 au and the tertiary (HOPS-92-B) is 499 au from the inner binary. The inner binary drives an outflow with a sharp u-shaped red lobe, and we cannot confidently detect a blue lobe. This outflow appears like it may be two overlapping outflows since there are two sets of u-shape edges visible, but ultimately we treat it as a single outflow and measure one PA. The outflow PA of the inner binary is 76$^\circ$ and the launching source is not distinguishable. The tertiary drives an outflow that is pointed almost 180$^\circ$ away from the previous outflow. It has a blue lobe with a sharp v-shape, and we cannot confidently detect a red lobe. The outflow PA of the tertiary is 269$^\circ$.

\cite{Federman_2023} looks at the ratio of flux densities probed by the ALMA Atacama Compact Array (ACA) and 12~m array at 0.1 arcsecond resolution which suggests a large envelope around the source. Thus, the classification of this source as flat spectrum is likely the result of viewing it close to face-on, meaning we are looking down the outflow cavity.

\subsubsection{HOPS-56}
HOPS-56 contains four sources. Sources A, B, and C are close-companions at similar distances from each other. Source D is much further away. We could not confidently detect nor measure any outflows in this field. In the blue channels, there is some filamentary and arcing structure near the three close-companions, but nothing that we can confidently identify as an organized outflow. In the red channels, there are two small shell-like structures launching from the close-companions, but the shapes are vague and we cannot confidently detect an outflow lobe. Source D looks like it has one edge of a v-shaped outflow pointing south, but we do not measure it due to a low signal-to-noise ratio and the lack of a second edge.

\subsubsection{HOPS-70}
HOPS-70 contains five sources, four of which form binary pairs. We could not confidently detect nor measure any outflows in this field. In the blue channels, both binary systems have arcing shapes that may be outflow-related. The red channels also contain some filamentary structures but no definite outflow shapes.

\cite{Federman_2023} looks at the ratio of flux densities probed by the ALMA ACA and 12~m array at 0.1 arcsecond resolution which suggests a large envelope around the source. Thus, the classification of this source as flat spectrum is likely the result of viewing it close to face-on, meaning we are looking down the outflow cavity.

\subsubsection{HOPS-84}
HOPS-84 contains two sources with a separation of 276 au. We measured one HC outflow with two lobes. The lobe-average PA is 83$^\circ$, the lobe PA offset is 3$^\circ$, and the launching source is not distinguishable. The blue lobe has a sharp v-shape and the red lobe has a faint outline of a v-shape, but its signal-to-noise ratio is quite low. It is possible there are two overlapping aligned outflows present, since in the blue lobe each source seems to feed into the outflow. Ultimately, due to ambiguity, we treat it as a single outflow and measure one PA. Additionally, there is a lot of emission inside the cavity, almost in a spiraling structure.

\subsubsection{HOPS-75}
HOPS-75 contains two sources with a separation of 100 au. We measured one LC outflow with one lobe. The outflow PA is 100$^\circ$ and the launching source is not distinguishable. The blue channels possess a diffuse u-shape structure that is significantly offset from the binary, so we cannot confidently measure an outflow PA. The red lobe has a diffuse v-shape, buried under other emission. There is lots of extraneous emission in the field. Due to the general messiness and ambiguity of the emission, we label these measurements LC.

\subsubsection{HOPS-85}
HOPS-85 contains two sources with a separation of 40 au. We could not confidently detect nor measure any outflows in this field. The blue channels have a ring/shell structure pointing away from the binary. The red channels have a similar structure but it is oriented 90$^\circ$ to the other one. While there are some directional structures, the overall situation is ambiguous.

\cite{Federman_2023} note that based on the appearance of the disk, we are viewing HOPS-85 nearly face-on, meaning we are looking down the outflow cavity.

\subsubsection{HOPS-77}
HOPS-77 contains three sources in a hierarchical triple system. We could not confidently detect nor measure any outflows in this field. The blue channels have no emission structure in the vicinity of the sources. The red channels contain a potential outflow lobe on the inner binary, but it is ambiguous.

\subsubsection{HOPS-182}
HOPS-182 contains three sources, of which two form a close-companion system with a separation of 382 au. We measured one HC outflow with two lobes. The lobe-average PA is 235$^\circ$, the lobe PA offset is 16$^\circ$, and the launching source is not distinguishable. In faster velocity channels the blue lobe is a diffuse u-shape launching from HOPS-182-A. In slower blue velocity channels, there appears to be diffuse u-shaped overlapping aligned outflows launching from both Sources A and B. The red lobe has a diffuse v-shape that in slower velocity channels appears to launch from HOPS-182-A, but in faster velocity channels it looks like both may be launching an outflow or feeding into the same one. Ultimately, we treat it as a single outflow and measure one PA. There is a third source much farther from the binary, HOPS-182-C, which appears to have a sharp v-shaped blue outflow lobe in velocity ranges from 3 to 5 km~s$^{-1}$, but we do not measure this outflow.

\subsubsection{HOPS-168}
HOPS-168 contains two sources with a separation of 74 au. We measured one HC outflow with two lobes. The lobe-averaged PA is 159$^\circ$, the lobe PA offset is 5$^\circ$, and the launching source is not distinguishable. Both lobes have sharp v-shapes.

\subsubsection{HOPS-203}
HOPS-203 contains three sources in a hierarchical triple system in which we measure two HC outflows, each with one lobe. The inner binary (HOPS-203-A and HOPS-203-B) has a separation of 50 au and the tertiary (HOPS-203-C) is 1086 au from the inner binary. The blue channels clearly display a sharp v-shape launching from the binary and a diffuse u-shape launching from the tertiary. In the red channels, two lobes are visible on each side of the binary but we do not measure these. The outflow PA for the inner binary is 144$^\circ$ and for the tertiary is 101$^\circ$. The launching source on the inner binary is not distinguishable.

\subsubsection{HOPS-173}
HOPS-173 contains two sources with a separation of 103 au. We measured one LC outflow with one lobe. The outflow PA is 247$^\circ$ and the launching source is not distinguishable. We do not detect a blue lobe, although the blue channels contain some arcing u-shaped structure north of the binary, but, given the PA measured in the red channels, do not seem associated with the outflow. The red channels display a sharp and faint v-shape on both sides of the binary and we measure the PA of the southwest lobe. Due to the low signal-to-noise, we label these measurements LC.

\subsubsection{HOPS-193}
HOPS-193 contains two sources with a separation of 100 au. We measured one HC outflow with two lobes. The lobe-averaged PA is 94$^\circ$, the lobe PA offset is 8$^\circ$, and the launching source is not distinguishable. The blue lobe has a diffuse u-shape only visible in two or three channels. The red lobe has a sharp v-shape also only visible in two or three channels.

\subsubsection{HOPS-158}
HOPS-158 contains two sources with a separation of 46 au. We measured one HC outflow with two lobes. The lobe-averaged PA is 236$^\circ$, the lobe PA offset is 18$^\circ$, and the launching source is not distinguishable. This outflow is wider than most. The blue channels have a u-shaped lobe, the base of which coincides with the binary. The red lobe is similar and appears to overlap with the blue lobe and encompass the continuum sources, but this appearance is possible at face-on or intermediate inclinations.

\subsubsection{HOPS-138}
HOPS-138 contains two sources with a separation of 126 au. We could not confidently detect nor measure any outflows in this field. Neither the blue nor red channels have any notable emission or structure. The maximum intensity map (moment 8) suggests the presence of a shell structure around the binary. However, this feature does not appear in any individual velocity channels, and is a superposition of seemingly unrelated emission.

\subsubsection{HOPS-395}
HOPS-395 contains two sources with a separation of 65 au. We measured one HC outflow with two lobes. The lobe-averaged PA is 174$^\circ$ and the lobe PA offset is 14$^\circ$. The outflow is launching from HOPS-395-B. The blue channels predominantly show the north lobe on the binary, but the south lobe is also strongly visible in some blue channels. Both lobes have diffuse v-shapes and are extremely narrow, like a jet. Their directions are mirrored across the separation line, and not directly across. This appearance is likely due to an edge-on orientation.

\subsubsection{HOPS-288}
HOPS-288 contains three sources in a hierarchical triple system in which we measure two HC outflows, each with one lobe. The inner binary (HOPS-288-A-A and HOPS-288-A-B) has a separation of 61 au and the tertiary (HOPS-288-B) is 222 au from the inner binary. In the blue channels, the binary and tertiary each clearly launch their own sharp v-shaped outflows, almost side-by-side. Also in the blue channels is a lobe on the opposite side that is more aligned with the outflow lobe on the tertiary, but has a trail of emission launching from the binary. In the red channels, the tertiary has a sharp v-shaped lobe but on the same side as before. In slower red velocity channels, there is a sharp v-shaped lobe on the opposite side but it is not clear whether it only belongs to the tertiary. We only measure a blue lobe for the inner binary and its PA is 56$^\circ$. The launching source on the inner binary is not distinguishable. We also only measure a red lobe for the tertiary and its PA is 212$^\circ$. We only measure these two lobes since it is ambiguous which source the other lobes belong to.

\subsubsection{HOPS-290}
HOPS-290 contains two sources with a separation of 333 au. We measured two HC outflows each with two lobes. The first outflow is a typical v-shape with two lobes, and launches from HOPS-290-A. The blue lobe is sharp and the red lobe diffuse. The lobe-averaged PA is 294$^\circ$ and the lobe PA offset is 4$^\circ$. The second outflow launches from HOPS-290-B. The blue lobe is atypical in that the outflow curves in, looking like a shell. This outflow has another lobe that also appears faintly on the blue side that we record as the red lobe. The lobe averaged PA is 229$^\circ$ and the lobe PA offset is 3$^\circ$. In continuum, HOPS-290-B is elongated while HOPS-290-A is not, suggesting the disks are misaligned.

\subsubsection{HOPS-281}
HOPS-281 contains two sources with a separation of 312 au. We measured one HC outflow with two lobes. The lobe-averaged PA is 26$^\circ$, the lobe PA offset is 3$^\circ$, and the launching source is not distinguishable. The blue lobe has a sharp u-shape. The red lobe has a sharp v-shape.

\subsubsection{HOPS-282}
HOPS-282 contains two sources with a separation of 70 au. We measured one outflow with two lobes. The lobe-average PA is 183$^\circ$, the lobe PA offset is 4$^\circ$, and the launching source is not distinguishable. Both lobes appear in both red and blue channels and each has a sharp v-shape. We assign the north lobe as the blue lobe and the south lobe as the red lobe.

\subsubsection{HOPS-261}
HOPS-261 contains two sources. We could not confidently detect nor measure any outflows in this field. In both the low red and blue channels, there are very compact and indiscernible structures near the binary.

\subsubsection{HOPS-384}
HOPS-384 contains ten sources, of which two form a close-companion system with a separation of 49 au. Only the first four sources are marked on the map. We measured one LC outflow with one lobe. The outflow PA is 184$^\circ$ and the launching source is not distinguishable. The blue channels have some blobs and an arc of emission in front of the binary but we cannot confidently detect an outflow lobe, nor measure an outflow PA. The red lobe has a sharp v-shape with lots of filament structure in the cavity and along the edges. Due to the general messiness of the emission, we label these measurements LC.

\subsubsection{HOPS-304}
HOPS-304 contains two sources with a separation of 273 au. We measured one LC outflow with two lobes. The lobe-averaged PA is 332$^\circ$, the lobe PA offset is 30$^\circ$, and the launching source is not distinguishable. The blue lobe has a diffuse u-shape and the red lobe has a faint diffuse u-shape. The red channels have some extraneous emission in the northwest corner of the field. Due to the large lobe PA offset and general messiness in the emission, we label our measurements LC.

\subsubsection{HOPS-400}
HOPS-400 contains two sources with a separation of 184 au. We measured one HC outflow with two lobes. The lobe-averaged PA is 267$^\circ$ and the lobe PA offset is 19$^\circ$. The outflow is clearly launching from HOPS-400-A. Both lobes have a narrow and diffuse v-shape.

\subsubsection{HOPS-213}
HOPS-213 contains two sources with a separation of 144 au. We measured one LC outflow with two lobes. The lobe-averaged PA is 264$^\circ$, the lobe PA offset is 5$^\circ$, and the launching source is not distinguishable. The blue lobe has a sharp but faint v-shape only visible in two velocity channels. The red lobe has a faint u-shape, also only visible in two velocity channels. Due to the overall fuzziness of the image and low signal-to-noise, we label these measurements as LC.

\subsubsection{HOPS-312}
HOPS-312 contains two sources with a separation of 135 au. We measured one HC outflow with two lobes. The lobe-averaged PA is 216$^\circ$ and the lobe PA offset is 1$^\circ$. Both lobes of the outflow have sharp v-shapes. The blue lobe launches primarily from Source A while the red lobe appears to launch primarily from Source B. In both instances, the other source appears to feed into the outflow to a much lesser extent. This indicates the presence of overlapping aligned outflows.

\subsubsection{HOPS-363}
HOPS-363 contains two sources with a separation of 80 au. We measured one LC outflow with two lobes. The lobe-averaged PA is 93$^\circ$, the lobe PA offset is 24$^\circ$, and the launching source is not distinguishable. The blue lobe has a very diffuse v-shape. The red lobe has a sharp v-shape. Both lobes are surrounded by extraneous emission, either part of the outflow or not, but the outflow shape is clearly there. Due to the large lobe PA offset and general messiness of the emission, we label these measurements as LC.

\subsubsection{HOPS-323}
HOPS-323 contains two sources with a separation of 274 au. We measured one HC outflow with two lobes. The lobe-averaged PA is 303$^\circ$, the lobe PA offset is 7$^\circ$, and the launching source is not distinguishable. The blue lobe exhibits two sharp and distinct v-shapes launching from each source, with a difference in PA of 5$^\circ$ to 10$^\circ$. The red lobe, however, exhibits one sharp v-shape, launching from between the two sources. It appears possible that there are two outflows, one from each source. Ultimately, we treat it as a single outflow and measure one PA due to ambiguity.

\subsubsection{HOPS-366}
HOPS-366 contains two sources with a separation of 61 au. We measured one HC outflow with one lobe. The outflow PA is 137$^\circ$ and the launching source is not distinguishable. In the faster blue velocity channels, there is some circular and shell-like structure aligned with the outflow that appears in the slower blue channels. The blue lobe has a sharp u-shape. The red side of the outflow is extremely crowded; there is generally outflow structure with the expected PA (opposite the blue lobe) but it is ambiguous so we cannot confidently measure an outflow PA.

\subsubsection{HOPS-361}
HOPS-361 contains 13 sources, of which two form a close-companion system (a = 38 au). Only the first four sources are marked on the map. The sources of interest are the binary consisting of HOPS-361-C-A and HOPS-361-C-B (collectively also known as IRS3). We measured one HC outflow with one lobe. The outflow PA is 201$^\circ$ and the launching source is not distinguishable. The blue lobe has a sharp and faint v-shape. Nearby in the field are HOPS-361-A (IRS1) and HOPS-361-B (VLA1). Around Sources A and B there are some circular shell-like structures in the blue channels. We cannot confidently detect a red lobe, although the red channels may have a v-shape with the expected PA (opposite the blue lobe) but it is buried under the emission around HOPS-361-A and HOPS-361-B.

\cite{Cheng&Tobin+2022} identify a clear disk appearance at 0.87 mm and 1.3 mm in the close binary HOPS-361-C (IRS3) and measure a disk radius of 103 au and an inclination of $\sim$67$^\circ$. They note that HOPS-361-C-A (IRS3A) is coincident with the geometric center of the disk and drives a radio jet with a position angle $\sim$15$^\circ$, which differs from our measurement by 6$^\circ$ (when flipped 180$^\circ$). \cite{Cheng&Tobin+2022} also identify a single-lobe bubble-like outflow around HOPS-361-A.

\subsubsection{HOPS-364}
HOPS-364 contains two sources with a separation of 66 au. We measured one HC outflow with two lobes. The lobe-averaged PA is 276$^\circ$, the lobe PA offset is 18$^\circ$, and the launching source is not distinguishable. Both lobes are sharp v-shapes.

\subsubsection{HH270VLA1}
HH270VLA1 contains two sources with a separation of 99 au. We measured one HC outflow with two lobes. The lobe-averaged PA is 50$^\circ$, the lobe PA offset is 1$^\circ$, and the launching source is not distinguishable. Both lobes have sharp v-shapes and both lobes are visible in both the red and blue channels. The blue channels predominantly feature the southwest lobe and the red channels predominantly feature the northeast lobe. This indicates an edge-on orientation.

\subsection{Fields Without Outflow Emission} \label{sec:appendix:fields-without-emission}
The following systems contain no evidence of any outflow structure whatsoever in $^{12}$CO ($J=2\rightarrow1$) moment maps: HOPS-28, HOPS-163, HOPS-242, HOPS-255, HOPS-248, and HOPS-357. Although all of these sources were detected in the 1.3 mm dust continuum, we did not identify or measure any outflows in these fields. Note that HOPS-357 has regions of emission in the maximum intensity map which is right around the systemic velocity, likely just emission from the envelope.

\begin{figure*}[ht!]
\centering
\includegraphics[width=0.86\textwidth]{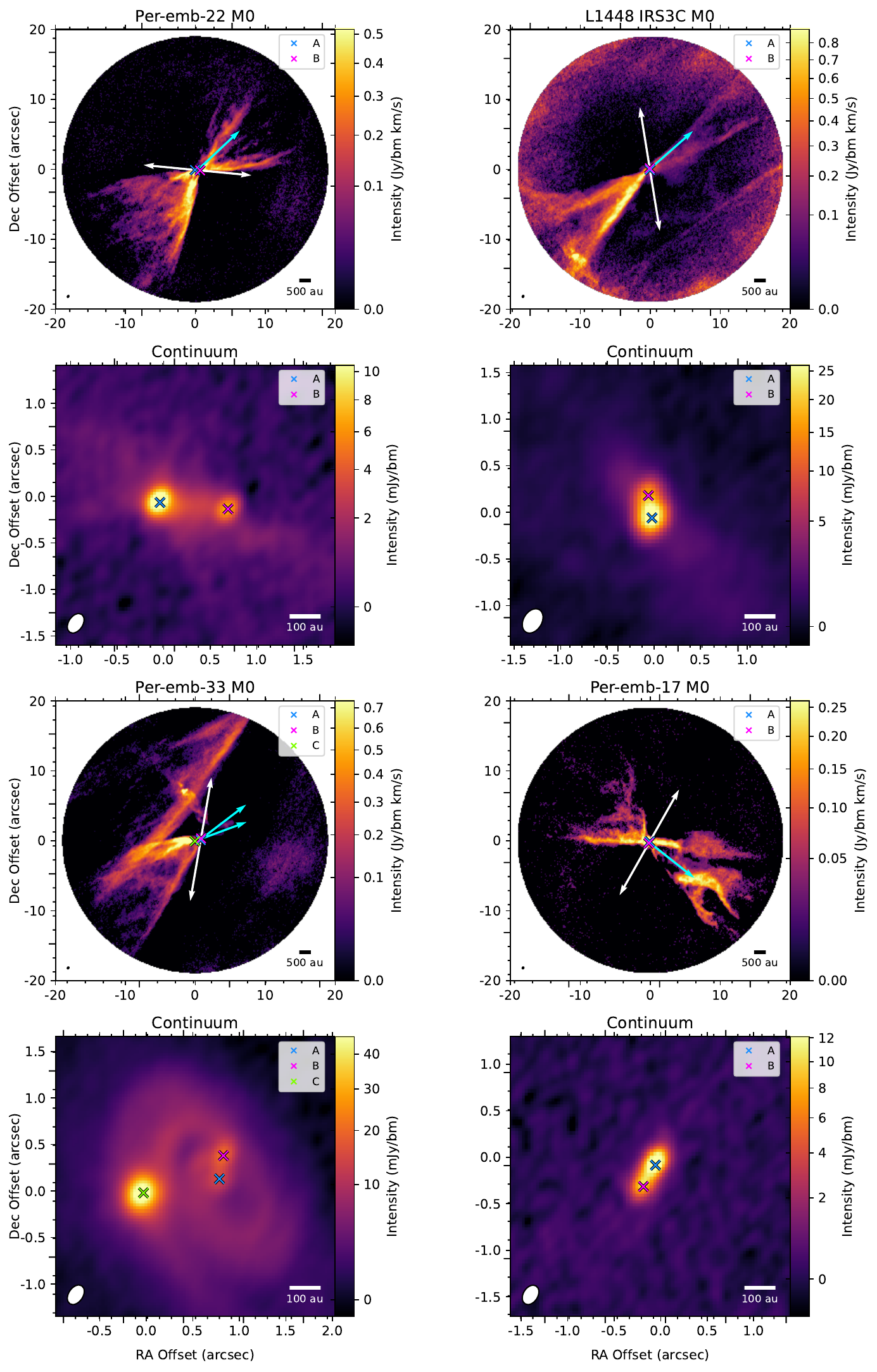}
\caption{All Fields (1 of 13). Same as Figure \ref{fig:fig_1}, but for all fields. Fields with no measured outflows are displayed with maximum intensity (moment 8) maps, as labeled in the map title.}
\label{fig:appendix-1}
\end{figure*}

\begin{figure*}[ht!]
\centering
\includegraphics[width=0.86\textwidth]{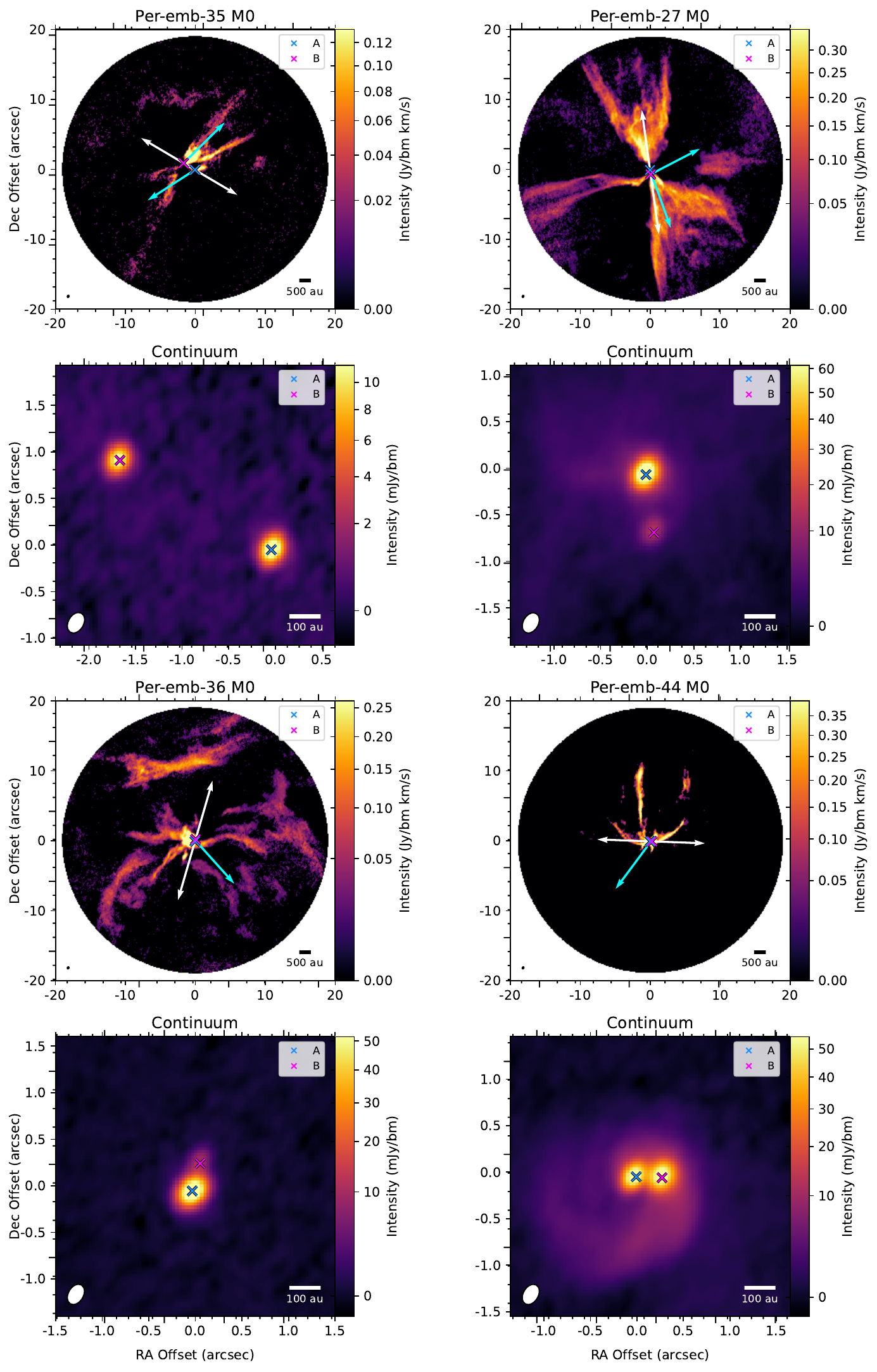}
\caption{All Fields (2 of 13). Same as Figure \ref{fig:fig_1}, but for all fields. Fields with no measured outflows are displayed with maximum intensity (moment 8) maps, as labeled in the map title.}
\label{fig:appendix-2}
\end{figure*}

\begin{figure*}[ht!]
\centering
\includegraphics[width=0.86\textwidth]{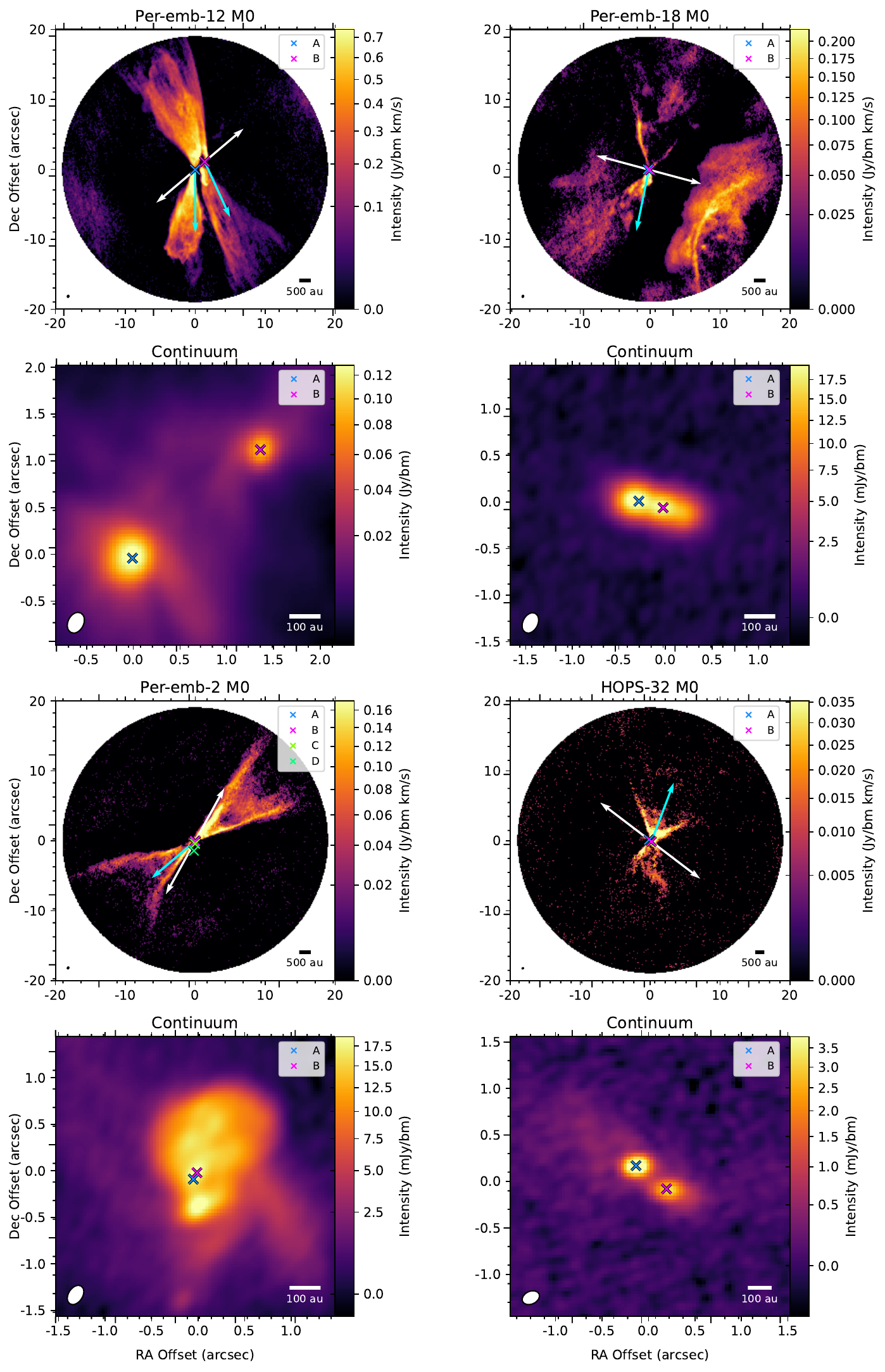}
\caption{All Fields (3 of 13). Same as Figure \ref{fig:fig_1}, but for all fields. Fields with no measured outflows are displayed with maximum intensity (moment 8) maps, as labeled in the map title.}
\label{fig:appendix-3}
\end{figure*}

\begin{figure*}[ht!]
\centering
\includegraphics[width=0.86\textwidth]{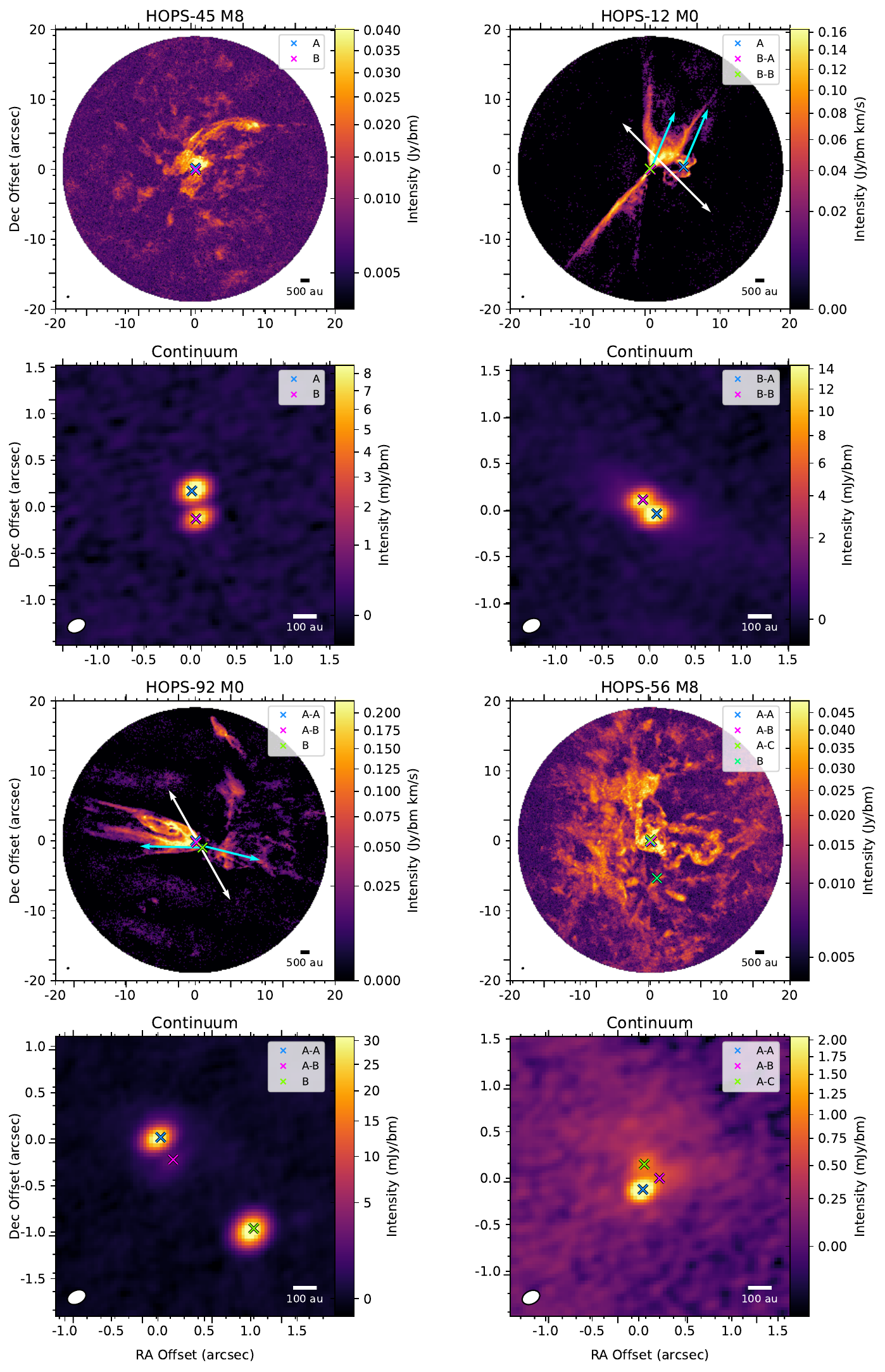}
\caption{All Fields (4 of 13). Same as Figure \ref{fig:fig_1}, but for all fields. Fields with no measured outflows are displayed with maximum intensity (moment 8) maps, as labeled in the map title.}
\label{fig:appendix-4}
\end{figure*}

\begin{figure*}[ht!]
\centering
\includegraphics[width=0.86\textwidth]{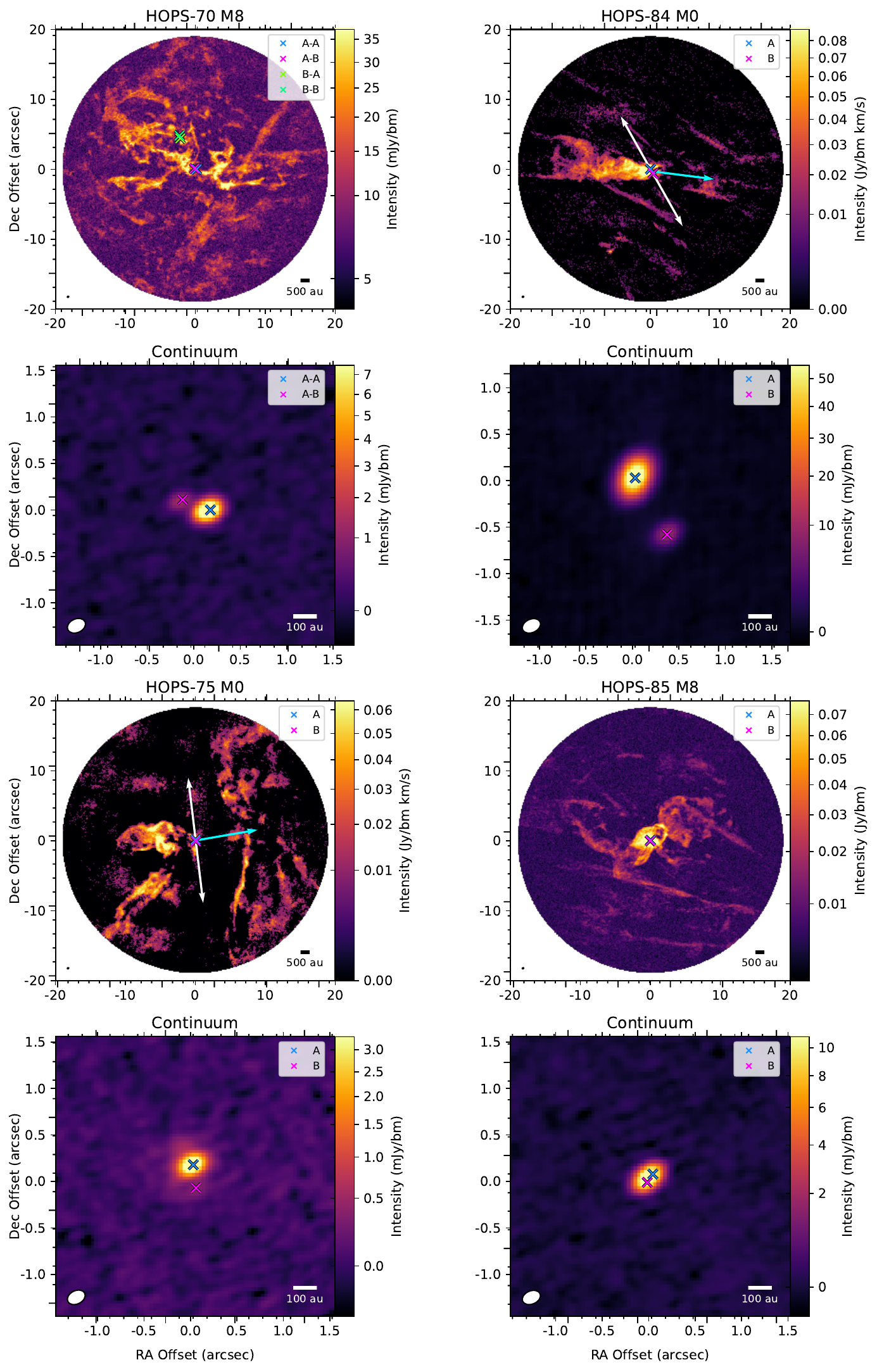}
\caption{All Fields (5 of 13). Same as Figure \ref{fig:fig_1}, but for all fields. Fields with no measured outflows are displayed with maximum intensity (moment 8) maps, as labeled in the map title.}
\label{fig:appendix-5}
\end{figure*}

\begin{figure*}[ht!]
\centering
\includegraphics[width=0.86\textwidth]{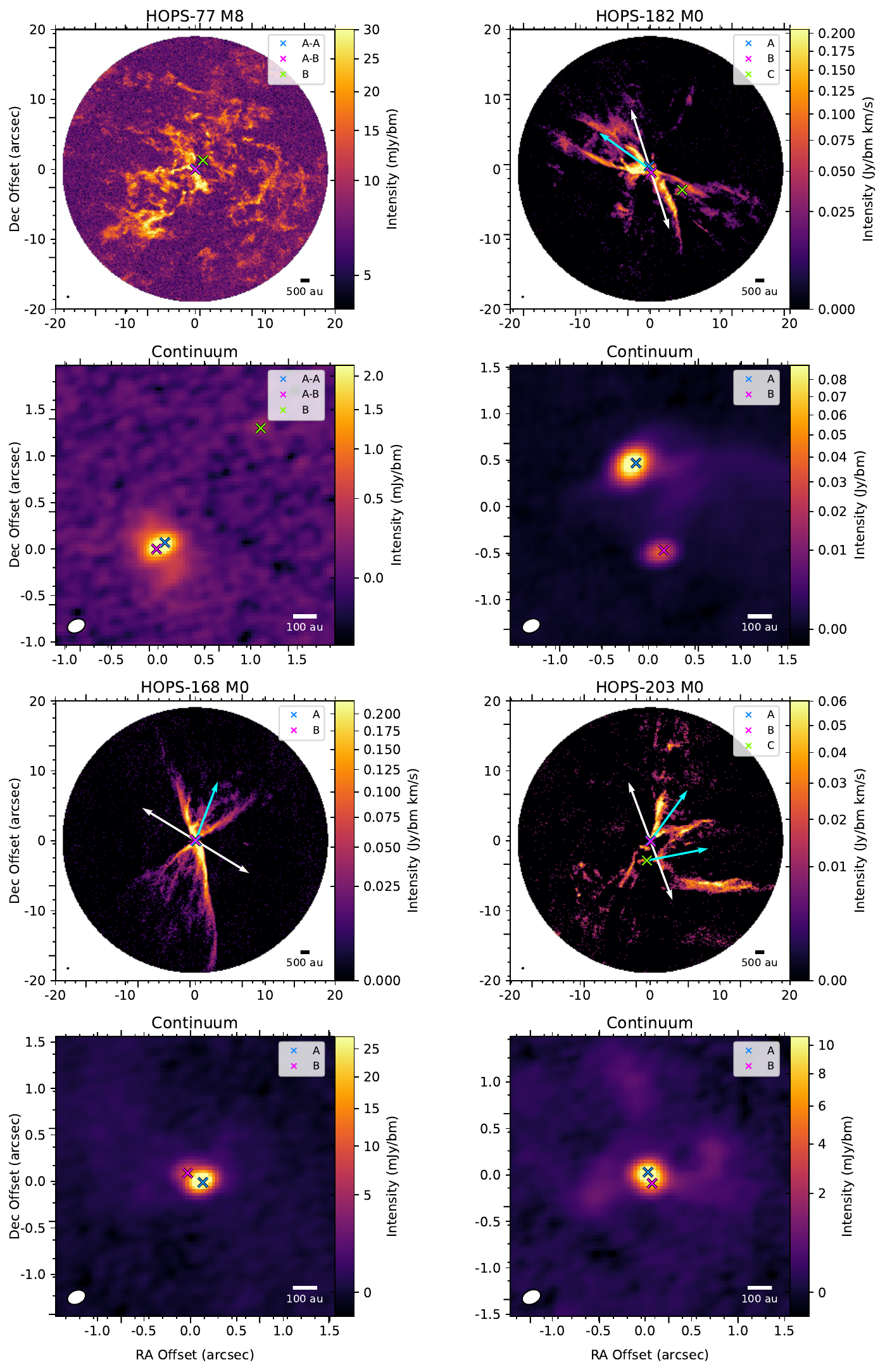}
\caption{All Fields (6 of 13). Same as Figure \ref{fig:fig_1}, but for all fields. Fields with no measured outflows are displayed with maximum intensity (moment 8) maps, as labeled in the map title.}
\label{fig:appendix-6}
\end{figure*}

\begin{figure*}[ht!]
\centering
\includegraphics[width=0.86\textwidth]{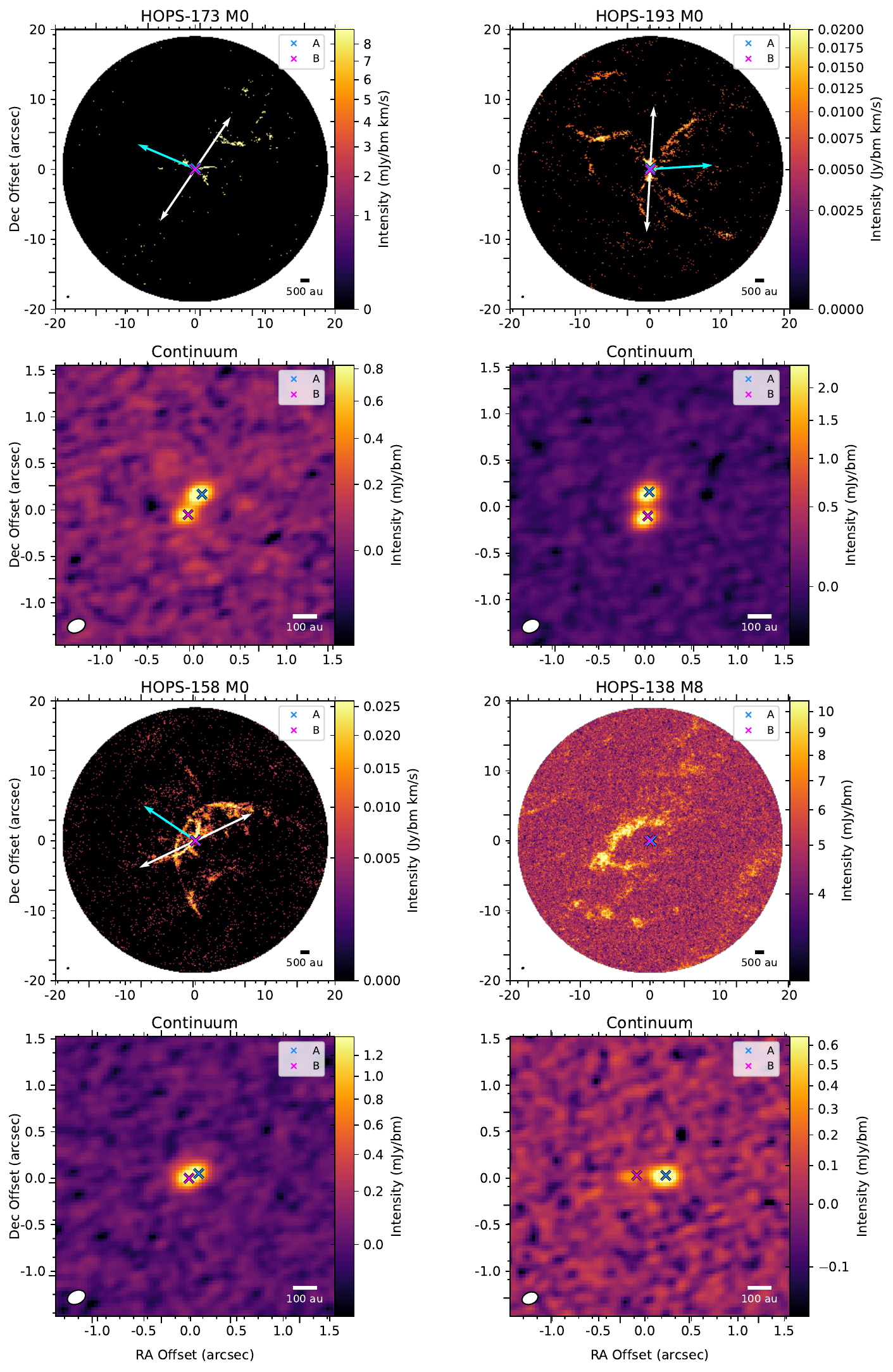}
\caption{All Fields (7 of 13). Same as Figure \ref{fig:fig_1}, but for all fields. Fields with no measured outflows are displayed with maximum intensity (moment 8) maps, as labeled in the map title.}
\label{fig:appendix-7}
\end{figure*}

\begin{figure*}[ht!]
\centering
\includegraphics[width=0.86\textwidth]{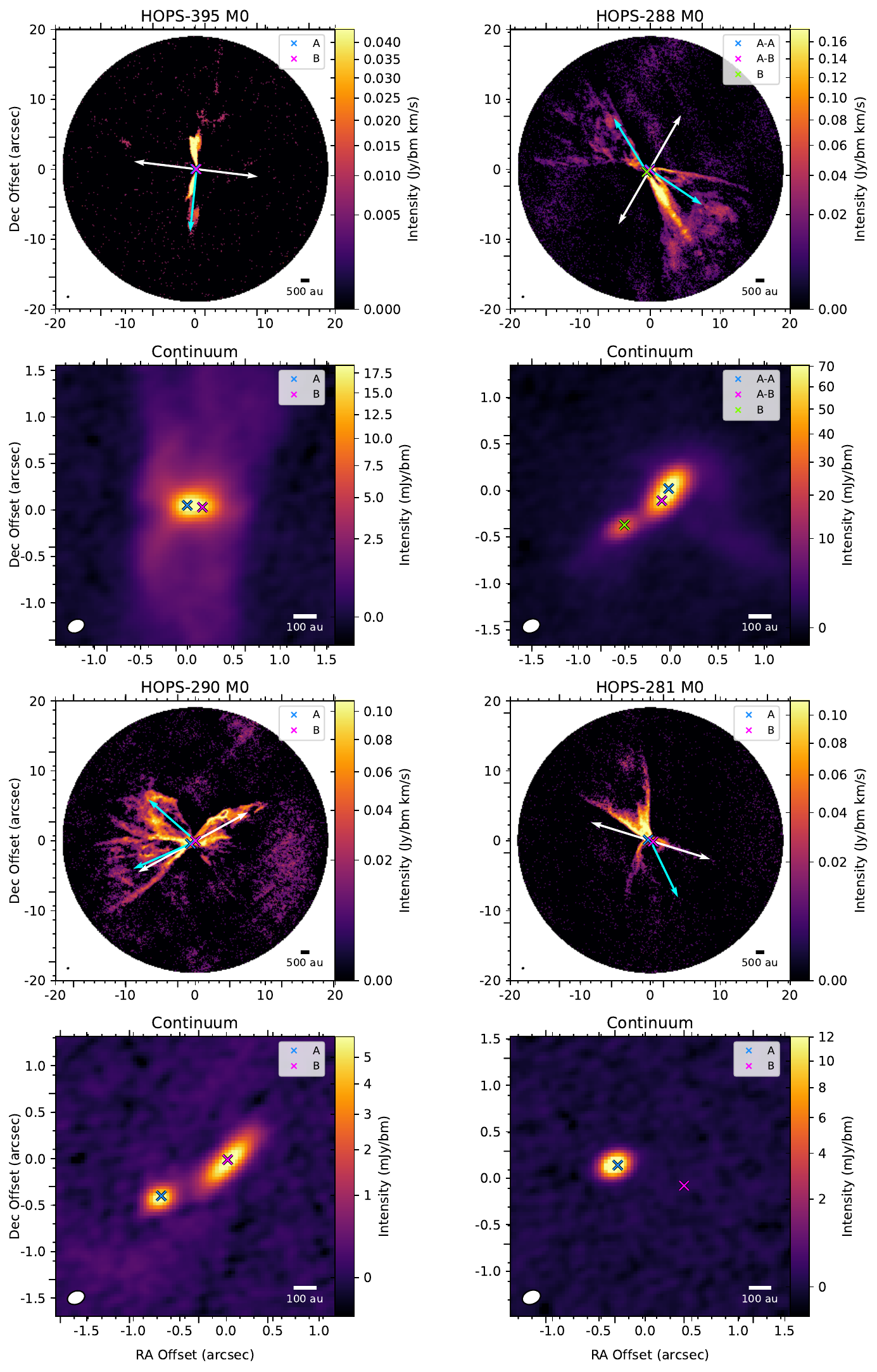}
\caption{All Fields (8 of 13). Same as Figure \ref{fig:fig_1}, but for all fields. Fields with no measured outflows are displayed with maximum intensity (moment 8) maps, as labeled in the map title.}
\label{fig:appendix-8}
\end{figure*}

\begin{figure*}[ht!]
\centering
\includegraphics[width=0.86\textwidth]{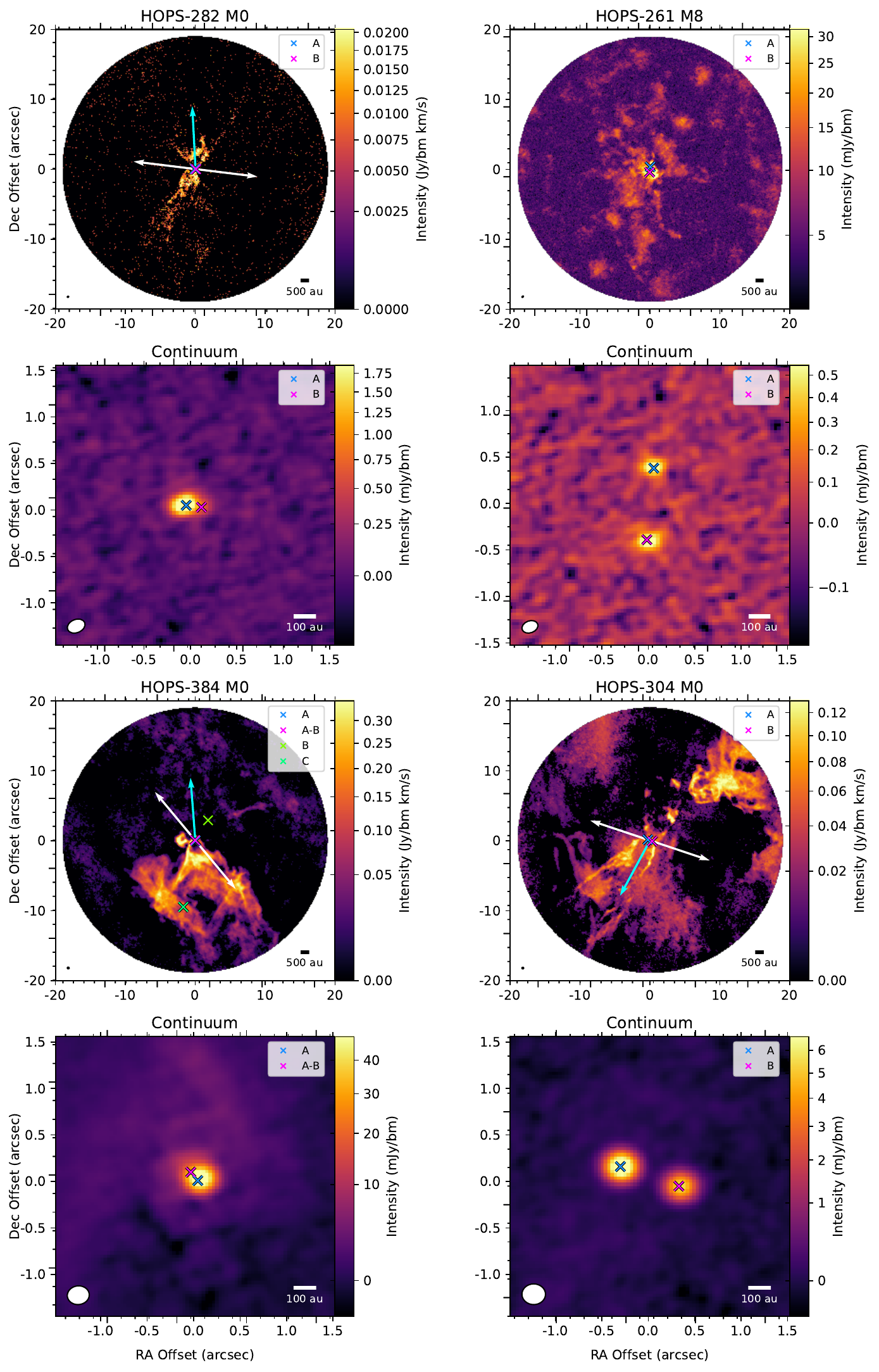}
\caption{All Fields (9 of 13). Same as Figure \ref{fig:fig_1}, but for all fields. Fields with no measured outflows are displayed with maximum intensity (moment 8) maps, as labeled in the map title.}
\label{fig:appendix-9}
\end{figure*}

\begin{figure*}[ht!]
\centering
\includegraphics[width=0.86\textwidth]{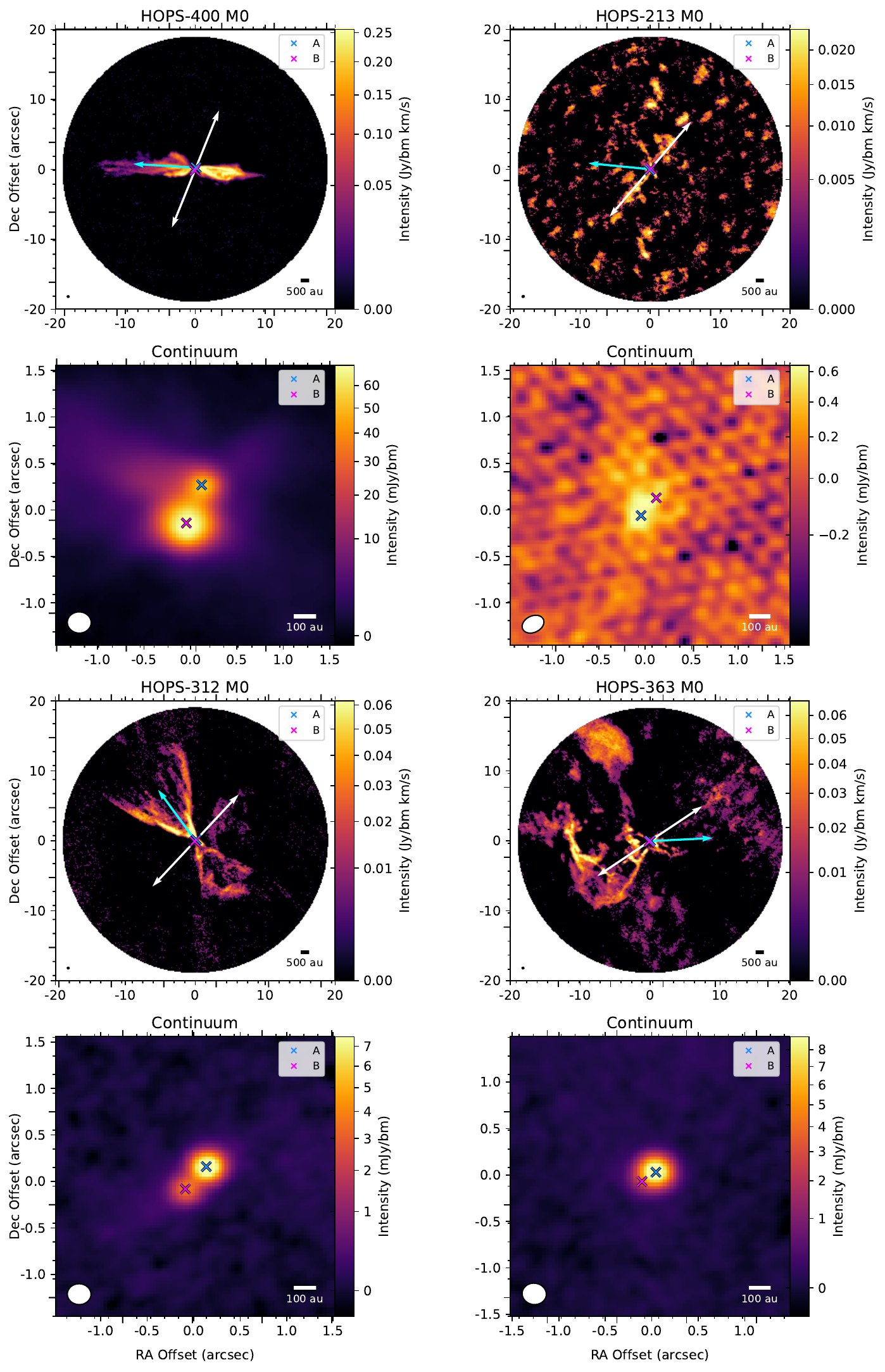}
\caption{All Fields (10 of 13). Same as Figure \ref{fig:fig_1}, but for all fields. Fields with no measured outflows are displayed with maximum intensity (moment 8) maps, as labeled in the map title.}
\label{fig:appendix-10}
\end{figure*}

\begin{figure*}[ht!]
\centering
\includegraphics[width=0.86\textwidth]{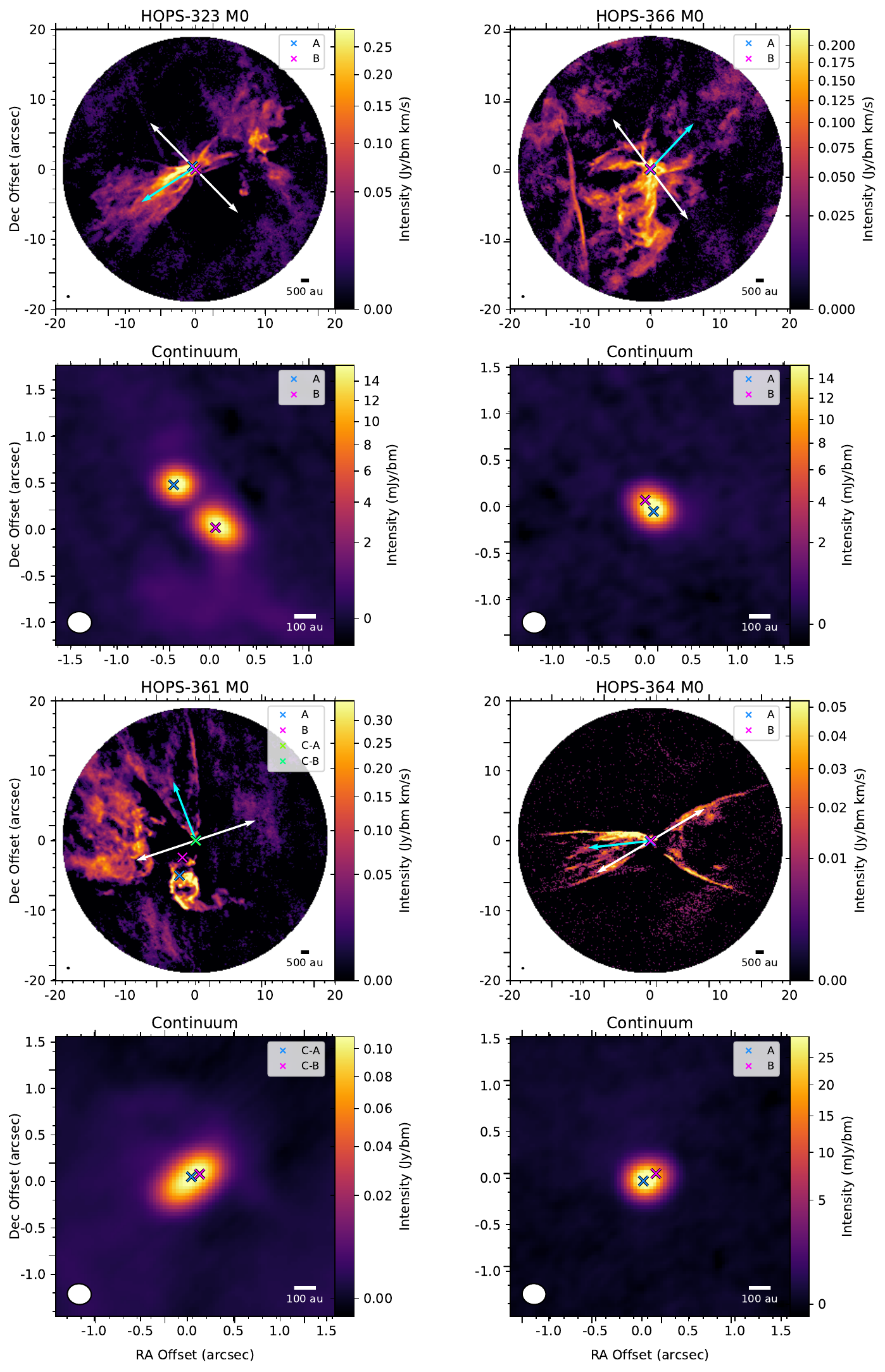}
\caption{All Fields (11 of 13). Same as Figure \ref{fig:fig_1}, but for all fields. Fields with no measured outflows are displayed with maximum intensity (moment 8) maps, as labeled in the map title.}
\label{fig:appendix-11}
\end{figure*}

\begin{figure*}[ht!]
\centering
\includegraphics[width=0.86\textwidth]{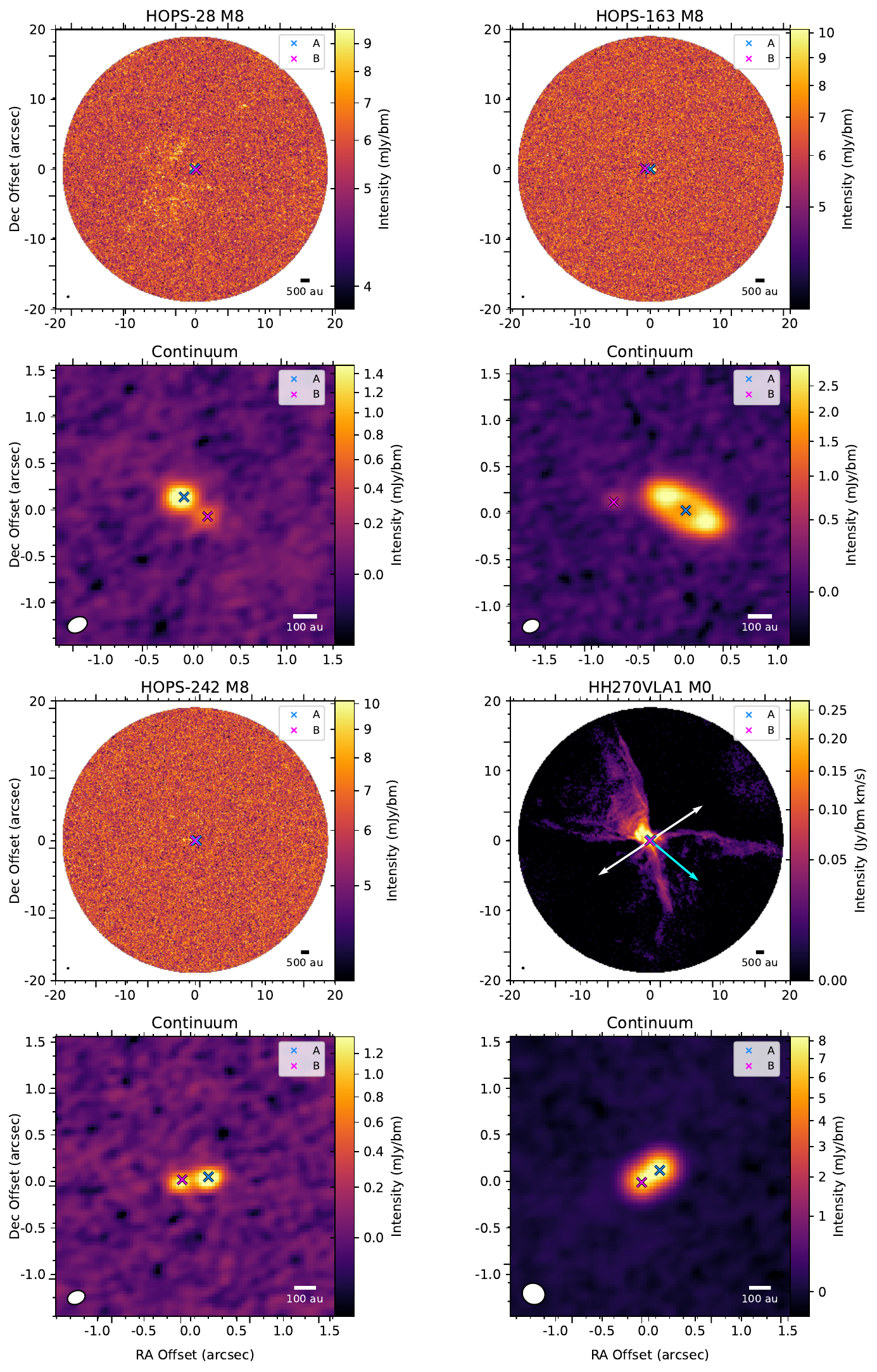}
\caption{All Fields (12 of 13). Same as Figure \ref{fig:fig_1}, but for all fields. Fields with no measured outflows are displayed with maximum intensity (moment 8) maps, as labeled in the map title.}
\label{fig:appendix-12}
\end{figure*}

\begin{figure*}[ht!]
\centering
\includegraphics[width=0.86\textwidth]{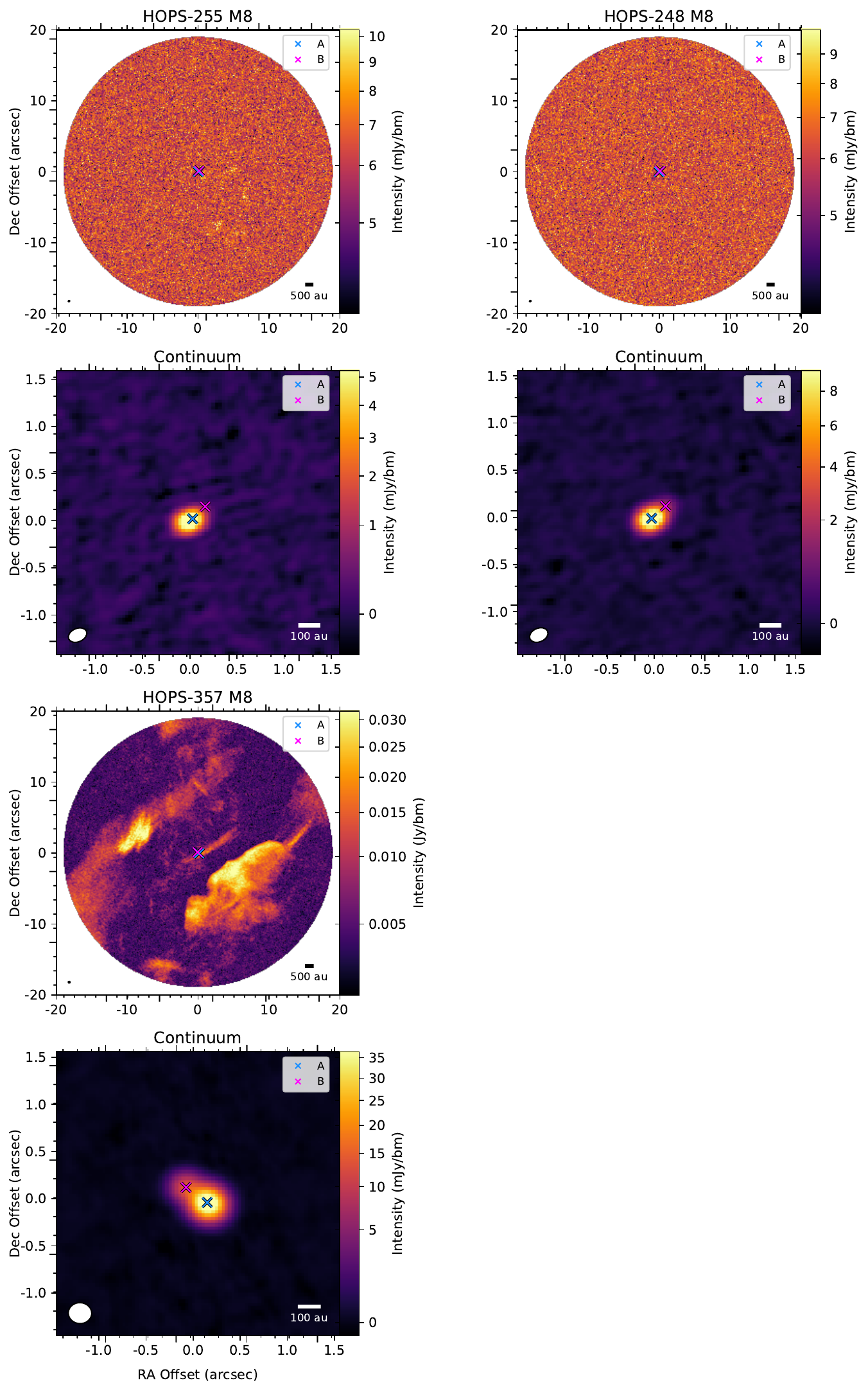}
\caption{All Fields (13 of 13). Same as Figure \ref{fig:fig_1}, but for all fields. Fields with no measured outflows are displayed with maximum intensity (moment 8) maps, as labeled in the map title.}
\label{fig:appendix-13}
\end{figure*}

%% This command is needed to show the entire author+affiliation list when
%% the collaboration and author truncation commands are used.  It has to
%% go at the end of the manuscript.
%\allauthors

%% Include this line if you are using the \added, \replaced, \deleted
%% commands to see a summary list of all changes at the end of the article.
%\listofchanges

\end{document}